\documentclass[11pt,letterpaper,twocolumn]{article}
\usepackage{usenix2019_v2}

\usepackage{silence}
\hypersetup{citecolor=blue,linkcolor=blue,draft}
\usepackage{amsfonts}
\usepackage{subcaption}
\usepackage{endnotes}
\usepackage{microtype}
\usepackage{graphicx}
\usepackage{fancyvrb}
\usepackage{multirow}
\usepackage{booktabs}
\usepackage{array,underscore,relsize}
\usepackage{fancyhdr}

\usepackage{enumitem}
\usepackage[labelfont=bf,font=small,skip=5pt]{caption}
\usepackage{xspace}
\usepackage{fp}
\usepackage{siunitx}

\usepackage[frozencache=true]{minted}

\usepackage{balance}

\usepackage{pifont}
\usepackage{wasysym}
\usepackage{cite}
\usepackage{algorithmic}
\usepackage{tikz}
\usepackage{textcomp}
\usepackage{bm}
\usepackage{xfrac}
\usepackage{wrapfig}
\usepackage{multicol}

\usepackage[compact,small]{titlesec}

\newcommand{\sys}{\mbox{\textsc{PACTight}}\xspace}

\newcommand{\CH}[1]{\textcolor{black}{#1}} 
\usepackage{soul,xcolor}
\setstcolor{red}

\newcommand{\yesmark}{\checkmark}
\newcommand{\nomark}{{\small $\times$}\xspace}

\input{glyphtounicode}
\newcolumntype{R}[1]{>{\raggedleft\let\newline\\\arraybackslash\hspace{0pt}}p{#1}}

\fvset{fontsize=\scriptsize,xleftmargin=8pt,numbers=left,numbersep=5pt}

\input{code/fmt}

\newcommand{\cc}[1]{\mbox{\smaller[0.5]\texttt{#1}}}

\ifdefined\THESIS

\else
\def\Snospace~{\S{}}

\fi

\newcommand{\x}{$\times$\xspace}

\newif\ifdraft\drafttrue
\newif\ifnotes\notestrue
\ifdraft\else\notesfalse\fi

\newcommand{\includepdf}[1]{
  \includegraphics[width=\columnwidth]{#1}
}

\newcommand{\squishlist}{
\begin{itemize}[noitemsep,nolistsep]
  \setlength{\itemsep}{-0pt}
}
\newcommand{\squishend}{
  \end{itemize}
}

\newcommand{\squishlists}{
\begin{itemize}[leftmargin=3mm,noitemsep,nolistsep]
  \setlength{\itemsep}{-0pt}
}
\newcommand{\squishends}{
  \end{itemize}
}

\newcommand{\nullsquishlist}{
\begin{description}[noitemsep,nolistsep]
  \setlength{\itemsep}{-0pt}
}
\newcommand{\nullsquishend}{
  \end{description}
}

\usepackage{tikz}

\usepackage{xstring}
\newcommand{\pp}[1]{
\vspace{2bp}
\noindent{\bf #1}
}

\newcommand{\etal}{\emph{et al.}\xspace}

\newcommand{\ie}{\emph{i.e.}}
\newcommand{\eg}{\emph{e.g.}}

\newcommand{\boxbeg}{
\par
\vspace{0.5em}
\noindent
\begin{tabular}{|l|}\hline
\begin{minipage}{3.2in}
\vspace{3px}
\noindent
}

\newcommand{\boxend}{
\vspace{3px}
\end{minipage}\\ \hline
\end{tabular}
\vspace{0.2em}
}

\ifdefined\THESIS
\newcommand{\tsection}[1]{\chapter{#1}}
\newcommand{\tsubsection}[1]{\section{#1}}

\else
\newcommand{\tsection}[1]{\section{#1}}
\newcommand{\tsubsection}[1]{\subsection{#1}}

\fi

\newcommand{\pacia}{\cc{pacia}\xspace}
\newcommand{\pacib}{\cc{pacib}\xspace}
\newcommand{\pacda}{\cc{pacda}\xspace}

\newcommand{\autia}{\cc{autia}\xspace}
\newcommand{\autib}{\cc{autib}\xspace}
\newcommand{\autda}{\cc{autda}\xspace}

\definecolor{applegreen}{rgb}{0.55, 0.71, 0.0}
\definecolor{cadmiumgreen}{rgb}{0.0, 0.42, 0.24}

\newcommand\spacebelowcaption{\vspace{-1.5em}}

\def\BibTeX{{\rm B\kern-.05em{\sc i\kern-.025em b}\kern-.08em
    T\kern-.1667em\lower.7ex\hbox{E}\kern-.125emX}}

\newcommand{\pctaddtag}{\mbox{\cc{pct\_add\_tag}}\xspace}
\newcommand{\pctrmtag}{\mbox{\cc{pct\_rm\_tag}}\xspace}
\newcommand{\pctsign}{\mbox{\cc{pct\_sign}}\xspace}
\newcommand{\pctauth}{\mbox{\cc{pct\_auth}}\xspace}

\begin{document}

\title{Tightly Seal Your Sensitive Pointers with \sys}


\newcommand{\osu}{{\textsuperscript{$\dagger$}}}

\author{
        Mohannad Ismail \quad
	Andrew Quach \osu \quad
	Christopher Jelesnianski \quad
	Yeongjin Jang \osu \quad
	Changwoo Min \quad \\
	Virginia Tech \quad \quad
	{\osu} Oregon State University
}


\maketitle

\begin{abstract}

\CH{ARM is becoming more popular in desktops and data centers, opening a new
realm in terms of security attacks against ARM.} 
ARM has released
\emph{Pointer Authentication}, a new hardware security feature that is intended
to ensure pointer integrity with cryptographic primitives.

In this paper, we utilize Pointer Authentication (PA) to build a novel scheme to
completely prevent any misuse of security-sensitive pointers. We propose \sys
to tightly seal these pointers. 
\sys utilizes a strong and
unique modifier that addresses the current issues with \CH{the state-of-the-art
PA defense mechanisms.} 
We implement four
defenses based on the \sys mechanism. 
Our security and performance evaluation results show that \sys defenses are more
efficient and secure. 
Using
real PA instructions, we evaluated \sys on 30 different applications,
including NGINX web server, with an average performance 
overhead of
\CH{4.07\%} 
even when enforcing our strongest defense.
\CH{\sys demonstrates its effectiveness and efficiency with real PA instructions 
on real hardware.}

\end{abstract}

\tsection{Introduction}
\label{s:intro}

In recent years, the ARM processor architcture started penetrating into the
data center~\cite{amazon:arm, oracle:arm, qualcomm:arm} and mainstream
desktop~\cite{apple:m1} markets beyond the mobile/embedded segments.
%
%
%
%
This opens a new realm in terms of security attacks against ARM, increasing the
importance of having effective and efficient defense mechanisms for ARM.

Control-flow hijacking attacks are one of the most critical security attacks.
These attacks aim to subvert the control-flow of a program by carefully
corrupting code pointers, such as return addresses and function pointers.
Control-flow integrity (CFI)~\cite{cfi} aims to defend against these attacks by
ensuring that the program follows its proper control-flow.
This is mainly done by generating a control-flow graph (CFG) of the program and
making the program conform to it.
%

In order to defend against control-flow hijacking attacks efficiently, ARM has
introduced a new hardware security feature, \emph{Pointer Authentication
(PA)}~\cite{qualcomm:pac}, which ensures pointer integrity with cryptographic
primitives. PA computes a cryptographic MAC called a \emph{Pointer
Authentication Code (PAC)} and stores it in the unused upper bits of a 64-bit
pointer. PA can be used to defend against control-flow hijacking attacks
securely and efficiently with low performance and memory overhead.

However, PA is not almighty. Although several PA-based defense mechanisms have
been proposed~\cite{parts-lilijestrand-sec19, pacstack-lilijestrand-sec21,
pcan-lilijestrand-systex19, ptauth-farkhani-sec21} and
deployed~\cite{apple:pac, qualcomm:pac}, we identified that they are still
exposed to attacks, such as using a signing gadget to forge
PACs~\cite{google:pac} and reusing PACs~\cite{parts-lilijestrand-sec19},
allowing arbitrary code execution.

In this paper, we propose \sys, which is a PA-based defense against
control-flow hijacking attacks.
In particular, we define three security properties of a pointer such that, if
achieved, prevent pointers from being tampered with. They are:
1) \emph{unforgeability}: A pointer \cc{p} should always point to its legitimate
object; 
2) \emph{non-copyability}: A pointer \cc{p} can only be used when it is at its
specific legitimate location; 
3) \emph{non-dangling}: A pointer \cc{p} cannot be used after it has been freed.
\sys tightly seals pointers and guarantees that a sealed pointer cannot be
forged, copied, or dangling.

\CH{Compared to previous PA-based defense mechanisms, \sys assumes a stronger
threat model such that an attacker has both arbitrary read and write
capabilities. \sys also provides better coverage by protecting a variety of
security-sensitive pointers. In this paper, we define a sensitive pointer as any
pointer that can reach a code pointer. \sys enforces the three properties
in order to prevent the pointers from being abused. Enforcement of the three
properties protects against attacks that rely on manipulating the pointers.}

\CH{We design \sys to achieve pointer integrity by protecting all sensitive
pointers and by providing spatial and temporal memory safety for those sensitive
pointers.}
Protecting these sensitive pointers achieves the balance between full memory
safety and covering only control-flow hijacking. This allows for reinforced
protection, thus achieving protection against control-flow hijacking attacks
and providing memory safety for sensitive pointers.
We demonstrate the effectiveness and practicality of \sys by evaluating with
real PA instructions on real hardware. 

In summary, we make the following contributions:
\squishlists

\item \CH{
We propose \sys, a novel and efficient approach to tightly seal pointers
using PAC.
By utilizing \sys's mechanisms, we make pointers unforgeable, non-copyable, and non-dangling.
}

\item We implemented four defenses using \sys: forward-edge protection,
backward-edge protection, C++ VTable pointer protection, and all sensitive
pointer protection.

\item We provide a strong security evaluation by demonstrating effectiveness
against real-world CVEs and synthesised attacks.

\item We evaluate \sys implementations on SPEC CPU2006, nbench, CoreMark
benchmarks, and NGINX web server with real PAC instructions. We show that \sys
implementations achieve low performance and memory overhead, \CH{4.07\%} and
23.2\% respectively making it possible to deploy \sys defenses in the
real-world.

\squishends

\section{Background and Motivation}
\label{s:bg}

%

In this section, we introduce control-flow hijacking attacks and ARM's pointer
authentication mechanism. We then discuss defenses based on PAC and their
limitations to motivate our work.

\subsection{Control-Flow Hijacking Attacks}
\label{s:bg:control}

%
Control-flow hijacking attacks are critical attacks to computer systems because
they may allow attackers to run arbitrary code on the system.
%
%
%
A popular way to carry out a control-flow hijacking attack is to exploit memory
corruption vulnerabilities, which C/C++ programs are prone to having.
In particular, attackers can alter the value of a code pointer (\eg, return
addresses and function pointers) by corrupting the memory location that stores
the pointer to subvert the execution flow of a program~\cite{bittau:brop,
jop-bletsch-asiaccs11, snow:jitrop, carlini:rop, davi:stitching,
cjujutsu-evans-ccs15, goktas:out}.

To defeat the attack, defenders must ensure that the program has no single
point that can let an attacker corrupt code pointers as well as data pointers that
refer to code pointers in its recursive memory dereference chain.
%
%
Return-oriented programming (ROP)~\cite{shacham:rop}, jump-oriented programming
(JOP)~\cite{jop-bletsch-asiaccs11}, and counterfeit-object oriented programming
(COOP)~\cite{schuster:coop} are the techniques that aim \CH{to achieve
code execution} by chaining returns, indirect call/jumps, and
virtual function calls in an object iteration loop, respectively.

\begingroup
\renewcommand{\arraystretch}{0.88} 
\begin{table*}[t]
        \centering
        \footnotesize
	\spacebelowcaption
	\resizebox{\textwidth}{!}{
\CH{
\begin{tabular}{l|lll|ccc}
\hline
\textbf{Defense} &
  \textbf{Protection scope} &
  \textbf{\begin{tabular}[c]{@{}l@{}}Attacker\\ abilities\end{tabular}} &
  \textbf{PAC modifier} &
  \textbf{\begin{tabular}[c]{@{}l@{}}Unforegablity\end{tabular}} &
  \textbf{\begin{tabular}[c]{@{}l@{}}Non-copyability\end{tabular}} &
  \textbf{\begin{tabular}[c]{@{}l@{}}Non-dangling\end{tabular}} \\ \hline
\multicolumn{1}{l|}{PARTS-CFI~\cite{parts-lilijestrand-sec19}} &
  \begin{tabular}[c]{@{}l@{}}Return addresses and \\ indirect code pointers\end{tabular} &
  \begin{tabular}[c]{@{}l@{}}Arbitrary read-write\end{tabular} &
  \begin{tabular}[c]{@{}l@{}}SP for return addresses \\ and type-id for indirect code pointers.\end{tabular} &
  \yesmark &
  \nomark &
  \nomark \\ [-2.5ex]
\multicolumn{1}{l|}{} &
   &
   &
   &
   &
   &
   \\ \hline
\multicolumn{1}{l|}{PACStack~\cite{pacstack-lilijestrand-sec21}} &
  Return addresses &
  \begin{tabular}[c]{@{}l@{}}Arbitrary read-write\end{tabular} &
  Previous return address on the stack &
  \yesmark &
  \yesmark &
  \nomark \\ [-2ex]
\multicolumn{1}{l|}{} &
   &
   &
   &
   &
   &
   \\ \hline
\multicolumn{1}{l|}{PTAuth~\cite{ptauth-farkhani-sec21}} &
  \begin{tabular}[c]{@{}l@{}}Heap allocated objects\end{tabular} &
  \begin{tabular}[c]{@{}l@{}}Arbitrary write\end{tabular} &
  A generated object-id &
  \yesmark &
  \nomark &
  \yesmark \\ [-2.2ex]
\multicolumn{1}{l|}{} &
   &
   &
   &
   &
   &
   \\ \hline
\multicolumn{1}{l|}{\textbf{\sys}} &
  \begin{tabular}[c]{@{}l@{}}All sensitive pointers \\ and return addresses\end{tabular} &
  \begin{tabular}[c]{@{}l@{}}Arbitrary\\ read-write\end{tabular} &
  \begin{tabular}[c]{@{}l@{}}
       The location of  the pointer and a random tag \\
       for sensitive pointers. Previous return address \\
       and a unique function id for return addresses. 
  \end{tabular} &
  \yesmark &
  \yesmark &
  \yesmark \\ \hline
\end{tabular}
}
}
	\caption{\CH{Comparison between \sys and state-of-the-art
    PAC-based defense mechanisms.}}
	\vspace{-20pt}
    \label{t:relwk}
\end{table*}
\endgroup

%
\subsection{ARM Pointer Authentication}
\label{s:bg:pac}

\begin{figure}[t]
	\centering
	\includegraphics[width=\columnwidth]{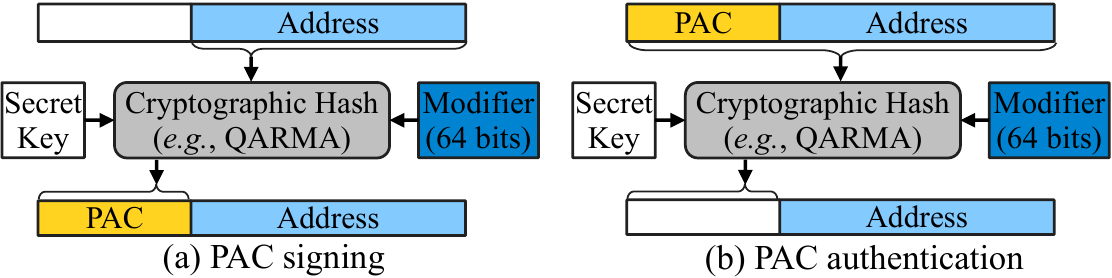}
	\spacebelowcaption
	\caption{
    PA signs a pointer and generates a pointer authentication code (PAC) based
    on a address, a secret key, a 64-bit user-provided modifier using PA
    instructions (\eg, \pacia). The signed pointer should be authenticated
    before the access using the same PAC, address, secret key, and modifier
    using PA instructions (\eg, \autia).
	}
	\vspace{-1.5em}
	\label{f:pac}
\end{figure}

%
%

ARMv8.3-A~\cite{qualcomm:pac} introduced a new hardware security feature,
Pointer Authentication (PA). PA has been implemented in the Apple A12 and M1
chips~\cite{apple:a12}.
%
The goal of PA is to protect the integrity of security-critical pointers, such
as code pointers.
%
%
%
To this end, a pointer authentication code (PAC) is generated by
a cryptographic hash function, as a message authentication code (MAC), to put
cryptographic integrity protection on the pointers.
A PAC is a MAC of the target pointer value, a secret key, and
a salt, which is a 64-bit modifier.
The modifier can be tweaked to bind the context of the program when generating
a PAC for a pointer.
Some examples of such context are conveying the type of the pointer as
a modifier, using stack frame address as a modifier, etc.

\pp{PAC signing.}
PAC utilizes a cryptographic hash algorithm, namely QARMA~\cite{qarma}. The 
algorithm takes two 64-bit values (pointer and modifier), as well as a 
128-bit key, and generates a 64-bit PAC.
These PACs are truncated and added to the upper unused bits of the 64-bit
pointer as illustrated in~\autoref{f:pac}(a). Five keys in total can be chosen to
generate the PACs.
These keys are stored in special hardware registers
protected by the kernel. 

\pp{PAC authentication.}
%
%
The cryptographic algorithm takes the pointer with the PAC and the modifier.
The PAC is then regenerated and compared with the one on the passed pointer.
To pass the authentication, both values need to be the same as the ones
originally used to generate the PAC.
If the regenerated PAC matches,
the PAC is removed from the pointer and the pointer can be used, as shown
in~\autoref{f:pac}(b).
Otherwise, the top two bits of the pointer are flipped, rendering the pointer
unusable.
Any use of the pointer results in a segfault.

\pp{PAC instructions.}
PAC instructions 
start with either \cc{pac} or \cc{aut} followed by a character that identifies
whether it protects a code pointer, data pointer or generates a generic PAC.
This is then followed by another character that identifies which key is being
used. For example, the \cc{pacib} instruction generates a PAC for a code
pointer that uses the B-key. When authenticating this code pointer, the
authenticate instruction for the code pointer and B-key, \ie, \cc{autib}, must
be used to successfully authenticate. Without this, the pointer cannot be used
as its semantics are changed. 


\subsection{PAC Defense Approaches}
\label{s:bg:attack}




\pp{Return address focused.}
Qualcomm's return address signing mechanism~\cite{qualcomm:pac} protects return
addresses from stack memory corruption.
It utilizes the \cc{paciasp} and \cc{autiasp} instructions.
These are specialized instructions that sign the return address in the Link
Register (LR) using the Stack Pointer (SP) as the modifier and the A-key
to protect return addresses.
%
%

However, because this approach is susceptible to PAC re-use attacks
(see \autoref{s:bg:limit}), PARTS return address
protection~\cite{parts-lilijestrand-sec19} includes the SP with a function ID
as a modifier to harden the PAC scheme against re-use attacks.
Moreover, PACStack~\cite{pacstack-lilijestrand-sec21} extends the modifier by
chaining PACs to bind all previous return addresses in a call stack.
On the other hand, PCan~\cite{pcan-lilijestrand-systex19} relies on protecting
the stack with canaries generated with PAC using a modifier consisting of
a function ID and the least-significant 48 bits from SP.

\pp{Other code pointers.}
Apple extended its protection to cover other pointer types including function
pointers and C++ VTable pointers. However, it uses a zero modifier to protect
them.
PARTS~\cite{parts-lilijestrand-sec19} utilizes PAC to protect function
pointers, return addresses, and data pointers. 
It utilizes a type ID based on~\emph{LLVM ElementType} as the modifier for
signing function pointers and data pointers.

\pp{Temporal safety.}
PTAuth~\cite{ptauth-farkhani-sec21} enforces temporal memory safety using PAC.
PTAuth generates a new random ID at each memory allocation and utilizes it as
a modifier for generating a PAC. Because the corresponding random ID of
a pointer is cleared or updated when the pointer is being freed or allocated,
PTAuth detects the violations of temporal memory safety (\eg, use-after-free)
by maintaining it as a modifier to
check the liveness of a pointer at the time of authentication.
%

\subsection{Limitations of Current PAC Defenses}
\label{s:bg:limit}


\pp{Forging PAC.}
%
%
PAC relies on the security of the cryptographic hash, that is, attackers cannot
generate a valid PAC for a pointer, even if they have both the pointer address
value and the corresponding modifier.
However, 
a memory corruption vulnerability in PAC generation logic may serve as an
arbitrary PAC generator, allowing attackers to bypass the PAC
authentication~\cite{google:pac}.

%

\pp{Reusing valid PACs in a different context.}
A PAC generated for one context can be reused in a different context if two
contexts share the same modifier.
This applies not only to the case of using zero modifier, such as Apple's
virtual function table protection, but also to the case that shares the same
modifier across different contexts, such as Qualcomm/Apple stack protection.
%
%
An example case of the latter is to reuse the PAC generated for a valid return
address with a specific SP at a different return location that shared the same SP
(\eg, having multiple function calls in a function, a case for sharing the same
stack frame for all of its returns).
%
PARTS-CFI is also susceptible to this attack because the approach uses a static
modifier for the pointer, based on its LLVM ElementType.
Having two different pointers of the same type, such two pointers will share
the same modifier, and in such a case, attackers can reuse the PAC generated
for one in the context of using the other.
\CH{~\autoref{t:relwk} summarizes the comparison between \sys and other
existing state-of-the-art PAC defense mechanisms.}
%

\pp{Reusing dangling PACs.}
Attackers can reuse legitimately generated PACs, even after a pointer becomes
dangling.
%
This occurs if the modifier used for signing the pointer does not convey the
temporal state of the pointer.
In such a case, the PAC is still valid even after deallocation of the
memory referred to by the pointer, and thereby, attackers may reuse a valid PAC
for a different object that the PAC has signed.
In particular, there is no mechanism in PARTS or Apple's Clang to
dynamically check and confirm if the pointers that they protect are not
\emph{dangling}, thus they are susceptible to this attack.

\section{Threat Model and Assumptions}
\label{s:threat}

Our threat model assumes a powerful adversary with read and write capabilities
by exploiting input-controlled memory corruption errors in the program. The
attacker cannot inject or modify code due to Data Execution Prevention (DEP),
which is by default enabled in most modern operating
systems~\cite{lwn:dep,windows:dep}. Also, the attacker does not control higher
privilege levels. We assume that the hardware and kernel are trusted,
specifically that the PA secret keys are generated, managed and stored
securely. Attacks targeting the kernel and hardware, such as
Spectre~\cite{spectre-kocher-sp19}, and
data only attacks, which modify and leak non-control data, are out of scope.
Our assumptions are consistent with prior works~\cite{parts-lilijestrand-sec19,
pacstack-lilijestrand-sec21, pcan-lilijestrand-systex19,
camouflage-denis-dac20} with the exception of
PTAuth~\cite{ptauth-farkhani-sec21}, 
%
which only allows arbitrary write and not arbitrary read.

\section{\sys Design}
\label{s:design}

\begin{figure*}[t]
	\centering
	\includegraphics[width=0.95\textwidth]{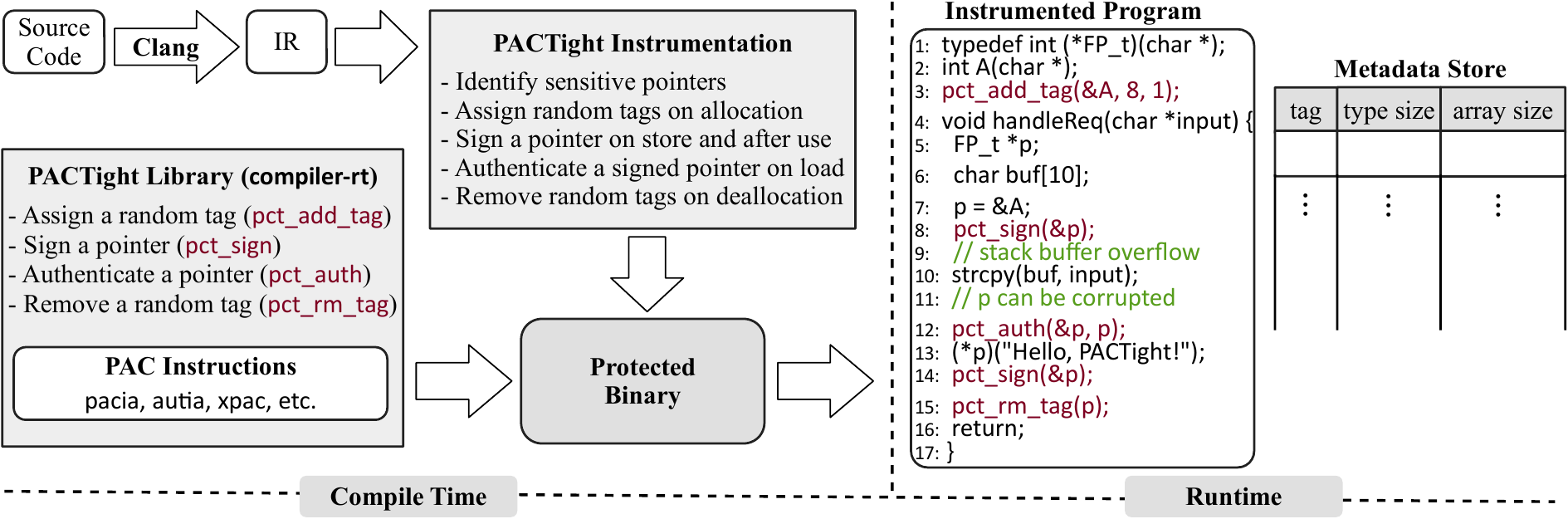}
    \caption{
    \sys design. At compile time, \sys instruments
    the allocation, assignment, use, and deallocation of code pointers and data
    pointers that are reachable to a code pointer (\ie, sensitive pointers).
    \sys guarantees three pointer integrity properties
    (\autoref{s:design:prop}), namely unforgeability, non-copyability, and
    non-dangling. At runtime, \sys generates a PAC for sensitive
    pointers using a novel authentication scheme and checks the PAC upon
    pointer dereference (\autoref{s:design:runtime}). \sys automates its
    instrumentation in four different levels: forward edge, backward edge,
    C++ VTable, and sensitive pointers (\autoref{s:defense}).
    }
	\vspace{-1.5em}
	\label{f:pactight-arch}
\end{figure*}

In this section, we describe the design of \sys. We first discuss our design
goal (\autoref{s:design:goal}), then we introduce three
pointer integrity properties that \sys enforces to overcome the limitations
of prior PAC approaches (\autoref{s:design:prop}), and then we compare \sys to current
state-of-the-art defenses (\autoref{s:design:compare}). Lastly, we present the
detailed design of \sys. As shown in \autoref{f:pactight-arch}, \sys consists
of a runtime library and compiler-based instrumentation. We first discuss the
runtime (\autoref{s:design:runtime}) to explain how \sys enforces the pointer
integrity properties and then explain \sys's automatic instrumentation and
defense mechanisms (\autoref{s:defense}).

\subsection{\sys Design Goals}
\label{s:design:goal}

The overarching goal of \sys is to completely prevent control-flow hijacking
attacks in a program with low performance overhead. While prior works on PAC
show promising results, they are limited in scope and/or security protection as
discussed in~\autoref{s:bg:attack}. To achieve our goal, it is essential to
enforce the complete integrity of pointers, which we will discuss
in~\autoref{s:design:prop}, and prevent any pointer misuse. We protect
sensitive pointers~\cite{kuznetsov:cpi} -- all code pointers and all data
pointers that are reachable to any code pointer -- because guaranteeing the
integrity of all sensitive pointers is sufficient to make control-flow
hijacking impossible. In summary, our main goals are:

\squishlists

\item \textbf{Integrity}: Prevent any misuse of sensitive pointers.

\item \textbf{Performance}: Minimize runtime performance and memory overhead.

\item \textbf{Compatibility}: Allow protection of legacy (C/C++) programs
without any modification.

\squishends

\subsection{\sys Pointer Integrity Property}
\label{s:design:prop}

\begin{figure}[t]
	\centering
	\includegraphics[width=\columnwidth]{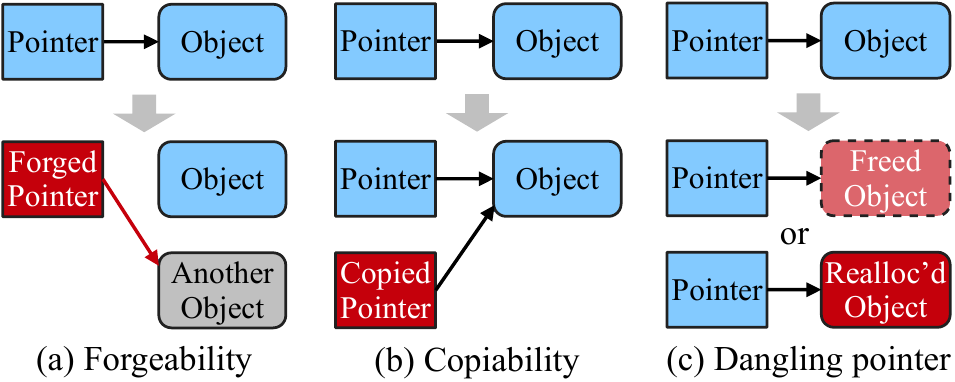}
	\spacebelowcaption

	\caption{
    Three types of violations of pointer integrity.
	}
	\vspace{-1.5em}
	\label{f:prop}
\end{figure}

Based on the limitations of prior PAC approaches and our observation on how
a pointer can be compromised, we define three security properties of pointer
integrity,
discussed in detail below:

\squishlists

\item \textbf{Unforgeability}: As illustrated in \autoref{f:prop}(a), a pointer
can be forged (\ie, corrupted) to point to an unintended memory object. Many
memory corruption-based control flow hijacking attacks fall into this category
by directly corrupting pointers (\eg, indirect call, return address). With the
\emph{unforgeability} property, a pointer always points to its legitimate
memory object and it cannot be altered maliciously.

\item \textbf{Non-copyability}: A pointer can be copied and re-used maliciously
as illustrated in \autoref{f:prop}(b). Many information leakage-based control
flow hijacking attacks first collect live code pointers and reuse the collected
live pointer by copying them to subvert control flow. With the
\emph{non-copyability} property, a pointer cannot be copied maliciously. It
asserts that a live pointer can only be referred from its correct location,
preventing the re-use of live pointers at different sites. \CH{If
\emph{non-copyability} is guaranteed, the security impact is
\emph{non-replayability}, and thus pointer attacks that replay PAC-ed pointers
for malicious use are prevented.}

\item \textbf{Non-dangling}: A pointer can refer to an unintended memory object if
its pointee object is freed or the freed memory is reallocated as shown in
\autoref{f:prop}(c). 
The integrity of a pointer is compromised even if the pointer itself is not
directly forged or copied. Semantically, the life cycle of a pointer should end
when its pointee object is destructed. Many attacks exploiting temporal memory
safety violation reuse such dangling pointers.
With the \emph{non-dangling} property, a pointer cannot be re-used after its
pointee object is freed. 

\squishends

The importance of these properties stems from the fact that to hijack
control-flow, at least one of these properties must be violated. \sys is able
to detect any of these violations before the use of a pointer, thus
guaranteeing the above mentioned pointer integrity. Note that ARM PAC only
enforces the unforgeability property.

\subsection{\CH{Comparison against Other PAC-based Defenses}}
\label{s:design:compare}


\CH{In contrast to other PAC-based defenses (\autoref{t:relwk}), \sys offers
more coverage against PAC attacks. PARTS~\cite{parts-lilijestrand-sec19} relies
on a static modifier based on the \emph{LLVM ElementType}, which can be
repeated. Even though an attack based on this would be harder than when using
the \cc{SP} as a modifier, it is still possible.
\sys's unique modifier scheme eliminates any \emph{replayability} of PACs, and
thus defends against PAC reuse. }

\CH{PACStack~\cite{pacstack-lilijestrand-sec21} introduces the idea of
cryptographically binding a return address to a particular control-flow path by
having all previous return addresses in the call stack influence the PA
modifier.
PACStack only protects return addresses on the stack and needs a forward-edge 
CFI scheme with it, whilst \sys protects all sensitive pointers on the stack 
and elsewhere.
}

\CH{PTAuth~\cite{ptauth-farkhani-sec21} attempts to provide protection against
temporal attacks. However, it assumes a weaker threat model, defending against
attackers with arbitrary write only. Also, it is vulnerable to intra-object
violation. 
If two pointers within the same object are swapped, PTAuth cannot detect this.
Thus, the pointers are \emph{copyable} in this case. Moreover, it only protects
the heap and does not handle stack protection, even from temporal attacks. 
\sys defends against a strong attacker, with arbitrary read and write
capabilities, protects the stack, heap, global variables, and defends against
any \emph{forging}, \emph{copyability}, and \emph{dangling} of pointers.}

%
%

\subsection{\sys Runtime}
\label{s:design:runtime}

This section describes the \sys runtime. \CH{We first describe how \sys efficiently
enforces the pointer integrity properties (\autoref{s:design:runtime:enforce}),
then discuss the \sys runtime library (\autoref{s:design:runtime:lib}), pointer 
operations (\autoref{s:design:runtime:pointer}) and the metadata store design 
(\autoref{s:design:runtime:meta}).}

\subsubsection{Enforcing \sys Pointer Integrity}
\label{s:design:runtime:enforce}

\begin{figure}[t]
	\centering
	\includegraphics[width=\columnwidth]{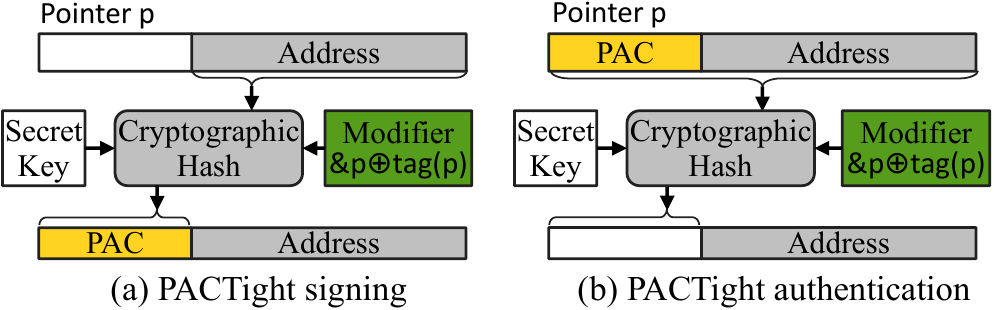}
	\spacebelowcaption
    \caption{
    Signing and authentication of a pointer variable \cc{p} in \sys. In
    addition to the unforgeability of \cc{p} provided by PA, \sys uses the
    address of a pointer (\cc{\&p}) and a random tag associated with a pointee
    (\cc{tag(p)}) to provide the non-copyability and non-dangling properties. 
    %
    }
	\vspace{-1.5em}
	\label{f:pactight-sign}
\end{figure}

In order to enforce the three properties, \sys relies on the PAC modifier.
The modifier is a user-defined salt that is incorporated by the cryptographic
hash into the PAC in addition to the address. Any changes in either the
modifier or the address result in a different PAC, detecting the violation. We
propose to blend the address of a pointer (\cc{\&p}) and a random tag
(\cc{tag(p)}) associated with a memory object to efficiently enforce the \sys
pointer integrity property, as illustrated in \autoref{f:pactight-sign}.

\squishlists

\item \textbf{Unforgeability}: PAC by itself enforces the unforgeability of
a pointer. PAC includes the pointer as one of the inputs to generate the PAC.
If the pointer is forged, it will be detected at authentication.

\item \textbf{Non-copyability}: \sys adds the location of the pointer
(\cc{\&p}) as a part of the modifier. This guarantees that the pointer can only
be used at that specific location. Any change in the location by copying the
pointer (\eg, \cc{q = p}) changes the modifier (\cc{\&q}) and thus triggers an
authentication fault.

\item \textbf{Non-dangling}: \sys uses a random tag ID to track the life cycle
of a memory object. \sys assigns a 64-bit random tag ID to a memory object upon
allocation and deletes it upon deallocation. This is done for both stack and
heap allocations. A random tag ID of a memory object (\cc{tag(p)}) is blended
with the location of the pointer (\cc{\&p}) to get the 64-bit modifier for PAC
generation and authentication. This implies that the life cycle of
a \sys-sealed pointer is bonded to that of a memory object. When memory is
deallocated (or re-allocated), \sys deletes (or re-generates) the random tag.
This invalidates all pointers to that memory, enforcing the non-dangling property. 

\squishends

By incorporating all these pieces of information (\ie, \cc{p}, \cc{\&p}, and
\cc{tag(p)}) together into the PAC, \sys effectively enforces the three
security properties for pointer integrity. Any change to any of the information
results in a PAC authentication failure. Note that we used XOR to blend the
location of a pointer and pointee's random tag into a single 64-bit integer.

\subsubsection{Runtime Library}
\label{s:design:runtime:lib}

The \sys runtime library provides four APIs to enforce pointer integrity. The
\sys LLVM instrumentation passes described in~\autoref{s:defense} automatically
instrument a program using those APIs. \CH{The code for this library is presented
in~\autoref{s:appendix:lib}.}

\pp{1) \cc{pct\_add\_tag(p,tsz,asz)}} sets the metadata for a newly allocated
memory region. Besides a pointer \cc{p}, it takes two additional arguments --
the size of an array element (\cc{tsz}) and the number of elements in the array
(\cc{asz}) in order to support an array of pointers. The \sys runtime assigns
\CH{the same random tag for each array element.} For each element, its
associated random tag and size information are added to the metadata store. 
\CH{This means that each array element's metadata can be looked up separately.} 
The API should be called whenever memory is allocated (heap or stack). 
\CH{\sys assigns a random tag to an object right after its allocation.}

\pp{2) \cc{pct\_sign(\&p)}} signs a pointer with the associated random tag that 
was generated by \pctaddtag. It generates a 64-bit modifier using the
location of a pointer (\cc{\&p}) and its associated random tag (\cc{tag(p)}) by
looking up the metadata store. Then, it signs the pointer with the modifier
using a PA signing instruction (\eg, \pacia, \pacda). If
a (compromised) program tries to sign a pointer that does not have an
associated random tag (\ie, the program tries to access unallocated memory
\CH{as in a use-after-free vulnerability}), \sys aborts the program. This API 
should be called whenever a pointer is assigned or after it is used.

\pp{3) \cc{pct\_auth(\&p\CH{,p+N})}} authenticates a pointer with the
associated metadata. Similar to \pctsign, it generates the modifier using the
pointer location (\cc{\&p}) and its associated random tag (\cc{tag(p\CH{+N})})
by looking up the metadata store, \CH{where \cc{N} is the array index. \cc{N}
is zero in cases other than arrays}. \CH{The use of \cc{p+N} allows support for
pointer arithmetic and enforcing spatial safety, which will be explained with
an example in~\autoref{s:design:runtime:pointer} (see~\autoref{c:array})}. Then,
it authenticates the pointer with the modifier using a PA authentication 
instruction (\eg, \autia, \autda). If there is no random tag 
or PA fails authentication, \sys aborts the program. \CH{Any value of
\cc{N} that is not within the bounds of the array will not return the correct
tag, and thus also causes a failed authentication.} If the authentication is
successful, it strips off the PAC from the pointer. This API should be called
before using the pointer. 

\pp{4) \cc{pct\_rm\_tag(p)}} removes the metadata associated to a pointer from
the metadata store. Once the metadata is deleted, any \pctauth to the deleted
memory will fail even if the memory is re-allocated. This API should be called
whenever memory (whether on the heap or the stack) is deallocated.

\subsubsection{Pointer Operations}
\label{s:design:runtime:pointer}

Since a \sys-signed pointer has a PAC in its upper bits, care must be taken to
not break the semantics of existing C/C++ pointer semantics. In particular, we
take care of the following four cases:

\pp{1) \sys-signed pointer comparison:}
Even if two pointers refer to the same memory address, their PACs are different
since the locations of the two pointers are different (\ie, \cc{\&p!=\&q}).
Hence, \sys strips off the PAC from the \sys-signed pointer before comparison
by looking for the \cc{icmp} instruction.

\pp{2) \sys-signed pointer assignment:}
When assigning one signed pointer (source) to another signed pointer (target),
the target pointer should be signed again with its location. 

\pp{3) \sys-signed pointer argument:} There are functions that directly
manipulate a pointer. For example, \cc{munmap} and \cc{free} take a pointer as
an argument and deallocate a virtual address segment or a memory block for
a given address. If their implementations do not consider PAC-signed pointers,
passing a PAC/\sys-signed pointer can cause segmentation fault. For those
functions, \sys strips off the PAC before passing the signed pointer as
an argument.

\begin{figure}[t!]
	\includegraphics[width=\columnwidth]{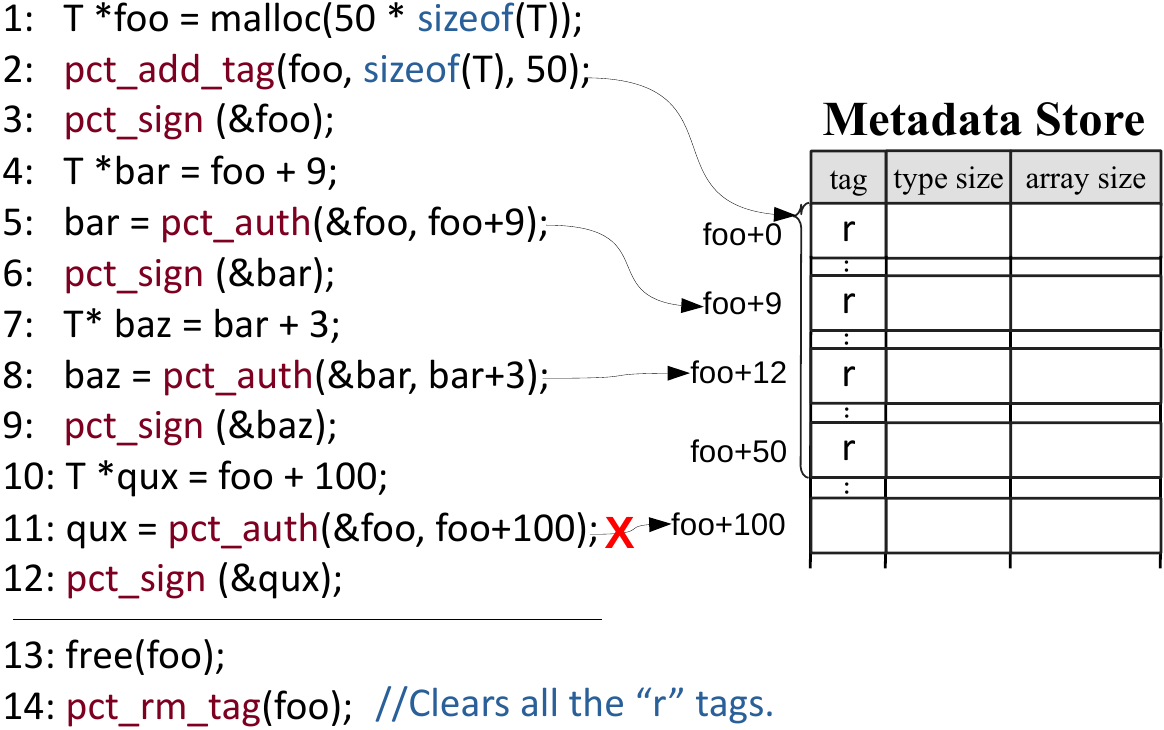}
	\spacebelowcaption

    \caption{
    \CH{Example of \sys handling array operations. 
    }
    }

	\vspace{-2.5em}
    \label{c:array}
\end{figure}

\pp{\CH{4) \sys-signed pointer arithmetic:}}
\CH{\sys supports pointer arithmetic on arrays. \sys assigns the same random
tag for all elements in an array, with the metadata keeping track of the size of
an element and the number of elements in the array to efficiently enforce
spatial safety. \autoref{c:array} shows a simplified representation of the metadata.
\sys first assigns the same random tag \cc{r} to all 50 array elements after the array 
allocation (Line 2). Each element has its own metadata. In Line 5, \sys successfully 
authenticates \cc{foo+9} using \cc{pct\_auth(\&foo,foo+9)}. \sys successfully 
authenticates \cc{bar+3} in Line 8, since \cc{bar+3} is \cc{foo+12}, and is within 
the array boundary. On the other hand, Line 10 violates spatial memory safety, and 
\sys throws an exception at Line 11. This is because \cc{tag(foo+100)} either does 
not exist or has a different tag. \autoref{c:libpac} shows the code of the runtime 
library.}
%

\CH{The mechanism works the same for temporal memory safety; a freed object
will not have a tag (Line 14), and newly allocated objects in the same location
will have a different tag. Thereby, \sys can effectively reject spatial and
temporal memory violations. Note that PTAuth~\cite{ptauth-farkhani-sec21}
performs ``backward search'' to find an array base address, which is not
necessary in \sys.}

\subsubsection{Metadata Store}
\label{s:design:runtime:meta}

\sys maintains a metadata store for allocated memory objects. For each
allocated memory object, the metadata store maintains a random tag, the size
of each individual element (or type size), and the number of elements in an
array (or array size). Non-array objects will be treated as an array having
a single element. \CH{We use either a 64-bit (default) or a 32-bit tag and we
compare the memory overhead between both tag sizes in~\autoref{s:eval:perf:bench}.}

We implemented the metadata store as a \CH{linear open addressing} hash table 
\CH{(base + offset)} using the address (\ie, \cc{p}) as the key. \CH{The base
address is kept in a reserved register, \cc{X18}, to avoid
leaking the metadata location (\ie, stack spill).} The metadata store is 
initialized when the program starts and is maintained by \sys's runtime 
library. An entry in the metadata store is allocated and deallocated using 
\pctaddtag and \pctrmtag, respectively. Whenever \sys needs to sign or 
authenticate (\pctsign, \pctauth), it looks up the metadata store to get 
the associated random tag and to check if the accessed memory is valid or not.
\CH{\sys relies on sparse address space support of the OS.
}

%

\subsubsection{A Running Example}
\label{s:design:runtime:example}

The code snippet in~\autoref{f:pactight-arch} (right) shows how \sys APIs are
used to protect a local function pointer \cc{p}. When \CH{an object} gets
allocated \CH{(Line 2)}, \pctaddtag allocates the metadata by setting a random 
tag and all the associated metadata. The number of elements and type size can 
be determined statically by analyzing the LLVM IR. Whenever a stack variable 
is assigned, the PAC is added with \pctsign \CH{(Lines 7, 8)}. 
If \CH{the pointer is dereferenced (Line 13) or if} any change in assignment
happens to the pointer legitimately, the PAC is authenticated (\pctauth)
\CH{(Line 12)} and a new PAC is generated for the new pointer with \pctsign
\CH{(Line 14)}. When a pointer gets deallocated \CH{(after the \cc{return} on
Line 16, since we are on the stack)}, the pointer is authenticated and all
metadata is removed (\pctrmtag). This is done by reading the type size and array
size from the metadata and removing the metadata accordingly. 

\section{\sys Defense Mechanisms}
\label{s:defense}

This section presents the \sys defense mechanisms built on top of the \sys
runtime. The \sys compiler passes automatically instrument all globals, stack
variables, and heap variables, inserting the necessary \sys APIs. We implement
four defense mechanisms: 1) Control-Flow Integrity (forward edge protection),
2) C++ VTable protection, 3) Code Pointer Integrity (all sensitive pointer
protection), and 4) return address protection (backward edge protection).

\subsection{Control Flow Integrity (\sys-CFI)}
\label{s:defense:cfi}

\sys-CFI guarantees forward-edge control-flow integrity by ensuring the \sys
pointer integrity properties for all code pointers. 
It authenticates the PAC on a function pointer at legitimate function call sites.
At all other sites, the code pointer is sealed with the \sys signing mechanism
so it cannot be abused. Any direct use of a \sys-signed pointer results in a segmentation 
fault, causing illegal memory access.

\pp{Instrumentation overview.}
In order to prevent any misuse and enforce all three security properties for
a code pointer, \sys-CFI should set metadata upon allocation and remove it upon
deallocation. Also, a function pointer should always be authenticated before
every legitimate use and it should be signed again afterwards. The \sys-CFI
instrumentation passes accurately identify and instrument all instructions in
the LLVM IR that allocate, write, use, and deallocate code pointers.

\pp{Identifying code pointers.} 
\sys-CFI identifies all code pointers using LLVM type information. Since code
pointers can be present inside composite types (\eg, \cc{struct} or an array of
\cc{struct}), \sys-CFI also recursively looks through all elements inside
a composite type. We specially handle the case that a code pointer is
manipulated after it is converted to some universal pointer type (\eg,
\cc{void*}). For example, for \cc{memcpy} and \cc{munmap} which take \cc{void*}
arguments, \sys-CFI gets the actual operand type first and instrumentation is
done accordingly. This is not only done for \cc{memcpy} and \cc{munmap}, but
for all universal pointer types. We look ahead for when they are typecasted
(\ie, \cc{BitCast} in LLVM IR) to get the original type accordingly

\pp{Instrumenting \sys APIs.}
Setting the metadata by instrumenting \pctaddtag is done immediately after all
code pointer allocations. This is done for all global, stack and heap
variables. In the case of initialized global variables, \pctaddtag and \pctsign
are appended to the global constructors. In this way, \sys-CFI maintains the
appropriate metadata for all global variables during program execution.

If the destination operand of the \cc{store} instruction is a code pointer,
\pctsign is instrumented right after the \cc{store} instruction to sign the
code pointer.

\pctauth must be called before any use of a code pointer. Specifically,
\sys-CFI looks for the \CH{relevant} \cc{load} and \cc{call} instructions 
and it instruments \pctauth immediately before the instructions. If the
authentication fails, the top two bits of the pointer are flipped meaning any
use of the pointer causes a segmentation fault, effectively denying any attack.
As the PAC authentication instructions (\eg, \autia) strips off the PAC, 
the PAC should be added again after the function call. Thus, \sys-CFI 
\CH{replaces the stripped pointer with the signed version} after indirect
\cc{call} instructions. This ensures that a PAC is always present. 

Whenever a code pointer is deallocated (\eg, \cc{free}, \cc{munmap}), \sys-CFI
removes the metadata by instrumenting \pctrmtag before the deallocation. For
stack variables, \pctrmtag is instrumented right before \cc{return}, and it
removes the metadata from the entire stack frame at once, from the first
variable to the last variable that has any metadata set. 

%
%

%
%
%

\pp{Summary.}
\sys-CFI is precise and efficient by enforcing the \sys pointer integrity
properties and leveraging hardware-based PA. Moreover, it provides the Unique
Code Target (UCT) property~\cite{hu:ucfi} because ensuring the \sys pointer
integrity properties implies that the equivalence class (EC) size (\ie, the
number of allowed legitimate targets at one call site) is always one. Thus, it
defends against all ConFIRM~\cite{li:confirm} attacks, which essentially rely
on the presence of more than one legitimate targets in an EC and replace an
indirect call/jump target with another allowed target. 

\subsection{C++ VTable Protection (\sys-VTable)}
\label{s:defense:vtable}

C++ relies on virtual functions to achieve dynamic polymorphism. At every
virtual function call, a proper function is used in accordance with the object
type. The mapping of an object type to a virtual function is done by the use of
a virtual function table (VTable) pointer, which is a pointer to an array of
virtual function pointers per object type. A VTable pointer is initialized in
the object's constructor and it is valid until an object is destructed.
Attacking the virtual function table pointer is a common exploit in C++
programs~\cite{burow:cfixx, schuster:coop, vtint-zhang-ndss15}.

\pp{Identifying VTable pointers.}
\sys-VTable identifies a VTable pointer in a C++ object by analyzing types in
LLVM. It investigates all composite types and checks if it is a class type
having one or more virtual functions. If so, it marks the first hidden member
of the class as a VTable pointer.
\CH{\sys-VTable also handles \cc{dynamic\_cast<T>}, since \cc{dynamic\_cast<T>} 
is only valid for a class with at least one virtual function pointer, so it 
has a virtual function table, and thereby they all are already considered 
sensitive types.}

\pp{Instrumenting \sys APIs.}
Upon a C++ type having a virtual function allocated, \sys-VTable instruments
\pctaddtag.
It instruments \pctsign immediately after the
VTable pointer is assigned by the object's constructor. This adds the PAC to the
pointer to seal it. Then, \pctauth is instrumented right 
before \cc{load}ing the VTable pointer. A failed authentication flips the top 
two bits of the pointer, rendering it unusable. 
Correspondingly, \pctrmtag is instrumented right before the object is destroyed 
(deallocation).

\begin{figure}[t!]
\begin{minted}{cpp}
\begin{Verbatim}[commandchars=\\\{\},codes={\catcode`\$=3\catcode`\^=7\catcode`\_=8}]
\PY{c+cm}{/** ==== nginx/http/ngx\PYZus{}http\PYZus{}variables.h ==== */}
\PY{k}{typedef} \PY{k}{struct} \PY{n}{ngx\PYZus{}http\PYZus{}variable\PYZus{}s} \PY{n}{ngx\PYZus{}http\PYZus{}variable\PYZus{}t}\PY{p}{;}

\PY{c+c1}{// a function pointer type (i.e., sensitive type)}
\PY{k}{typedef} \PY{n+nf}{ngx\PYZus{}int\PYZus{}t} \PY{p}{(}\PY{o}{*}\PY{n}{ngx\PYZus{}http\PYZus{}get\PYZus{}variable\PYZus{}pt}\PY{p}{)}\PY{p}{(}\PY{p}{.}\PY{p}{.}\PY{p}{.}\PY{p}{)}\PY{p}{;} \PY{err}{@}\PY{err}{\PYZbs{}}\PY{n}{label}\PY{p}{\PYZob{}}\PY{n+nl}{c}\PY{p}{:}\PY{n+nl}{nginx}\PY{p}{:}\PY{n}{fp2}\PY{p}{\PYZcb{}}\PY{err}{@}

\PY{k}{struct} \PY{n}{ngx\PYZus{}http\PYZus{}variable\PYZus{}s} \PY{p}{\PYZob{}} \PY{err}{@}\PY{err}{\PYZbs{}}\PY{n}{label}\PY{p}{\PYZob{}}\PY{n+nl}{c}\PY{p}{:}\PY{n+nl}{nginx}\PY{p}{:}\PY{n}{st}\PY{p}{\PYZcb{}}\PY{err}{@}
    \PY{n}{ngx\PYZus{}str\PYZus{}t}                 \PY{n}{name}\PY{p}{;}  
    \PY{c+c1}{// sensitive function pointer}
    \PY{n}{ngx\PYZus{}http\PYZus{}get\PYZus{}variable\PYZus{}pt}  \PY{n}{get\PYZus{}handler}\PY{p}{;} \PY{err}{@}\PY{err}{\PYZbs{}}\PY{n}{label}\PY{p}{\PYZob{}}\PY{n+nl}{c}\PY{p}{:}\PY{n+nl}{nginx}\PY{p}{:}\PY{n}{stfp2}\PY{p}{\PYZcb{}}\PY{err}{@}
    \PY{c+c1}{// ...}
\PY{p}{\PYZcb{}}\PY{p}{;} \PY{c+c1}{// a sensitive data type}
\end{Verbatim}
\end{minted}
	\spacebelowcaption
    \caption{
    Example of a sensitive data pointer in the (simplified) NGINX source code.
    %
    }
	\vspace{-1.5em}
	\label{c:sensitive}
\end{figure}

\subsection{Code Pointer Integrity (\sys-CPI)}
\label{s:defense:cpi}

\sys-CPI increases the coverage of \sys-CFI to guarantee integrity of all
sensitive pointers~\cite{kuznetsov:cpi}. Sensitive pointers are all code
pointers (\ie, \sys-CFI coverage) and all data pointers that point to code
pointers. It is possible to hijack control-flow by corrupting a sensitive
\emph{data} pointer because it can reach a code pointer. \autoref{c:sensitive}
shows an example of sensitive pointers from NGINX. A function pointer type
\cc{ngx\_http\_get\_variable\_pt} at Line~\ref{c:nginx:fp2} is a sensitive code
pointer. Also, a \cc{struct} type \cc{ngx\_http\_variable\_s} at
Line~\ref{c:nginx:st} is a sensitive data type because it has another sensitive
pointer (\cc{get\_handler} at Line~\ref{c:nginx:stfp2}) in it. If a sensitive
data pointer or its array index are corrupted, an attacker can hijack the
control-flow without directly corrupting the function pointer. 

\pp{Identifying sensitive pointers.}
\sys-CPI expands the type analysis of \sys-CFI to include all sensitive
pointers. It classifies a composite type that contains a function pointer as
a sensitive type. Then, it recursively classifies a composite type that
contains any sensitive pointer in it as a sensitive type until it cannot find
any more sensitive types. 
%
\CH{ We over-approximate when detecting security-sensitive pointers. 
That is, we regard a pointer as security-sensitive if we cannot
determine if it is non-security-sensitive statically (\eg, C union).
This approach may add extra instrumentation, however, it will not
compromise \sys’s security guarantees. 
}
\pp{Instrumenting \sys APIs.}
Instrumentation is then done in a similar manner to \sys-CFI by instrumenting
all instructions that allocate, store, modify and use sensitive pointers. In
case the pointers are of universal type (\ie, \cc{void*} or \cc{char*}),
\sys-CPI gets its actual type by looking ahead for a typecast and then
instrumentation is done accordingly. 

\subsection{Return Address Protection (\sys-RET)}
\label{s:defense:return}

Protecting return addresses is critical because they are, after all, the root
of ROP attacks. Meanwhile, the return address protection scheme should impose
minimal performance overhead because function call/return is very frequent
during program execution. We aim to minimize the signing/authentication
overhead without compromising the \sys pointer integrity properties. 

\pp{No non-dangling in return address.}
One interesting fact is that a return address cannot be a dangling
pointer.\footnote{Precisely speaking, a return address can be a dangling
pointer for Just-In-Time (JIT) compiled code in a managed runtime (\eg, Java,
Python). However, protecting control-flow hijacking in a managed runtime is the
out of scope for \sys. }
Hence, the non-dangling property doesn't need to be enforced and random tags are unnecessary.
Not using a tag offers
large performance benefits as metadata store lookup cost to get the random
tag can be removed.

\pp{Binding all previous return addresses.}
Instead of blending the location of a return address in a stack to provide the
\emph{non-copyability} property, we use the \emph{signed} return address of
a previous stack frame. Since the stack distance to a return address in
a previous stack frame is determined at compile time, accessing the previous
return address with a constant offset binds the current return address to the
relative offset of the previous stack frame (\ie, the current stack frame).
Hence we can achieve the \emph{non-copyability} property for return addresses.
In addition, by blending the signed return address of a previous stack frame,
we chain all previous return addresses to calculate the PAC of the current
return address. This approach is inspired by 
PACStack~\cite{pacstack-lilijestrand-sec21}.  \CH{Both \sys-RET and PACStack
incorporate the entire callstack in the modifier to prevent the reuse attack.}
\CH{In regards to dynamic stack allocation, the \cc{alloca()} function can 
dynamically adjust the stack frame size. 
To support dynamic stack allocation, \sys-RET uses LLVM intrinsics, such 
as \cc{getFrameInfo()} and \cc{getCalleeSavedInfo()}, that allow us to find 
the previous stack frame and 
calculate the distance correctly.}
%

\pp{Signing and authentication of a return address.}
Our optimized sign/authentication scheme for return addresses is as follows. We
blend a caller's unique function ID and the signed return address from the
previous stack frame to generate the modifier. This blending allows us to
achieve the \emph{non-copyability} property by chaining all previous return
addresses (binding a return address to a control-flow path), alongside the
guarantee of the \emph{unforgeability} property achieved by the PAC mechanism.
Instrumentation is done in the MachineIR level during frame lowering. Frame
lowering emits the function prologues and epilogues. The PAC is added at the
function prologue and authenticated at the function epilogue. The LLVM-assigned
function ID is unique due to the use of link time optimization (LTO). 
\CH{
\subsection{Optimization to Reduce PAC Instructions}
\label{s:defense:opt}
The main source of overhead in \sys would be due to the cryptographic
operations done by the QARMA algorithm. 
This is done every time a PAC instruction is executed. 
%
As discussed in \autoref{s:design:runtime:example}, \pctauth strips the PAC from the
pointer and \pctsign is added again after the pointer is used to add the PAC
again, thus maintaining the seal on the pointer. Thus, instead of re-adding the
PAC with \pctsign, we save the original pointer with the PAC before \pctauth in
a temporary register, and overwrite the stripped pointer with a PACed pointer
without needing to call \pctsign. Note that our code generation pass prevents
the stack spill of the temporary register to avoid the register from being
restored.
}

%
  %
  %
%

\tsection{Implementation}
\label{s:impl}


Our prototype consists of 4014 lines of code (LoC), with 3237 LoC for the
LLVM pass, 656 LoC for the \sys runtime library, and 121 LoC for the
AArch64 backend. \sys-CFI, \sys-VTable and \sys-CPI are all implemented in the
LLVM IR level while \sys-RET is implemented in the AArch64 backend. The \sys
runtime library is integrated with LLVM as part of \cc{compiler\_rt}. 
%
\CH{We use CSPRNG seeded by hardware RNG (RNDR in ARMv8)~\cite{arm:rndr} for
random tag generation. } 
To harden our prototype, we used different key types for sensitive
function pointers (\pacia, \autia), sensitive data pointers (\pacda, \autda),
and return addresses (\pacib, \autib).

\CH{We apply several optimizations to \sys. First, we use Link Time
Optimization (LTO), which combines all the object files into one file. Then, we
inline all our \sys runtime library functions. Finally, we implement the
additional optimization, discussed in~\autoref{s:defense:opt}, to reduce PAC
instructions. The evaluation of the impact of these optimizations is discussed
in \autoref{s:eval:perf:opt}.}
We also make our code for \sys public at
\url{https://github.com/cosmoss-jigu/pactight}.

\section{Evaluation}
\label{s:eval}

We evaluate \sys by answering the following: 

\squishlists

\item How effectively can \sys prevent not only synthetic attacks but also
real-world attacks by enforcing \sys pointer integrity properties? (\autoref{s:eval:sec})

\item How much performance and memory overhead does \sys impose?
(\autoref{s:eval:perf})

\squishends

\subsection{Evaluation Methodology}
\label{s:eval:method}

\pp{Evaluation environment.}
We ran all evaluations on Apple's M1 processor~\cite{apple:m1}, which is the
only commercially available processor supporting ARMv8.4 architecture with ARM
PA instructions. Specifically, we used an Apple Mac Mini M1~\cite{apple:mac}
equipped with 8GB DRAM, 4 big cores, and 4 small cores. We ported our prototype
to Apple's LLVM 10 fork~\cite{apple:llvm}. For all applications, we enabled
\cc{O2} and \cc{LTO} optimizations for fair comparison.

\pp{Evaluation of C applications.}
We ran all C applications with real ARM PA instructions. In this case, we
turned off all Apple LLVM's use of PA~\cite{apple:pac} to avoid the conflicting
use of PA instructions.

\pp{Evaluation of C++ applications.}
During initial evaluation, we found that the use of PA instructions is built into
Apple's standard C++ library. 
We have investigated 
using Ubuntu Linux~\cite{corellium:m1} on the M1 to work around this problem.
At time of writing, the Linux kernel on Ubuntu/M1 does not support PA --
the kernel does not activate PA during the boot procedure -- so userspace
applications cannot use PA instructions. 

For C++ applications, we use two different approaches to validate if \sys's
instrumentation is correct and to get an accurate performance estimation. 
For the correctness testing, we ran all C++ applications on ARM Fixed Virtual
Platform (FVP)~\cite{arm:fvp}, which is an ARM hardware platform simulator that
supports pointer authentication. We used the FVP only to test correctness, since it
is not a cycle-accurate simulator. We ran Linux on FVP to run C++ applications,
and we modified the Linux kernel and bootloader to activate ARM PA. All our C++
applications 
passed the correctness testing with FVP.
To simulate the overhead of a PA instruction and to get accurate performance
estimates on real hardware, we measured the time to execute a PA instruction
and found that seven XOR (\cc{eor}) instructions take almost the same time --
0.15\% faster -- to execute one PA instruction on the Apple Mac Mini M1.
Similarly, Lilijestrand \etal~\cite{parts-lilijestrand-sec19} also replaced
a PA instruction with four \cc{eor} instructions to estimate the performance
overhead, which is more optimistic than our measurement on hardware.

\subsection{Security Evaluation}
\label{s:eval:sec}

In this section, we evaluate \sys's effectiveness in stopping security attacks
using three real-world exploits (\autoref{s:eval:sec:cve}) and five synthesized
exploits (\autoref{s:eval:sec:syn}).

\subsubsection{Real-World Exploits}
\label{s:eval:sec:cve}

We evaluated \sys with three real-world exploits to test its effectiveness
against real vulnerabilities. 

\pp{(1) CVE 2015-8668.}
This is a heap-based buffer overflow~\cite{cve-2015-8668-exploit}
\CH{corrupting a sensitive pointer} in 
the \cc{libtiff} library. 
The heap overflow overwrites a function pointer in the \cc{TIFF} structure,
which allows attackers to achieve arbitrary code execution. \sys-CFI/CPI
successfully detects this and stops it from completing by enforcing \pctauth on
the corrupted function pointer.

\pp{(2) CVE-2019-7317.}
This is a use-after-free exploit~\cite{cve-2019-7317-exploit} 
in the \cc{libpng}~\cite{libpng} library. 
The \cc{png\_image\_free} function is called indirectly and frees
memory that is referenced by \cc{image}, \CH{a sensitive pointer}. \cc{image}
is then dereferenced. 
Since \sys-CPI does recursive identification, \cc{image} is instrumented.
When \cc{image} gets dereferenced after the free, \sys-CPI will detect that no
metadata exists 
and halts the execution.

\pp{(3) CVE-2014-1912.}
This is a buffer overflow vulnerability~\cite{cve-2014-1912-exploit} in
\cc{python2.7} that happens due to a missing buffer size check. An attacker can
corrupt a function pointer in the \cc{PyTypeObject} 
and achieve arbitrary code execution. \sys-CFI/CPI detects this by detecting the
corrupted function pointer with \pctauth.

\subsubsection{Synthesized Exploits}
\label{s:eval:sec:syn}

\pp{CFIXX test suite.}
We evaluated \sys with five synthesized attacks for C++ to demonstrate how
\sys-VTable can defend against virtual function pointer hijacking attacks, COOP
attacks \cite{schuster:coop} -- \CH{an attack} that crafts fake C++
objects. We used CFIXX C++ test suite~\cite{cfixx-test-suite-burow-web} by
Burow \etal~\cite{burow:cfixx}. It contains four virtual function pointer
hijacking exploits (\cc{FakeVT-sig}, \cc{VTxchg-hier}, \cc{FakeVT},
\cc{VTxchg}) and one COOP exploit. To make the test suite more similar to real
attacks, we modified the suite to use a heap-based overflow rather than
directly overwriting with \cc{memcpy}. This modification is similar to
a synthesized exploit in OS-CFI \cite{khandaker:oscfi}. 
\sys-VTable detects all the exploits by enforcing \pctauth on the virtual
function pointer before the virtual function call. 
The COOP attack crafts a fake object without calling the constructor
and utilizes a virtual function pointer of the fake object. \sys-VTable detects
this due to the fact that it was never initialized and thus \pctauth fails.

\pp{\CH{Vulnerable code to other PAC defenses.}}
\CH{ We describe here a synthesized exploit that bypasses
PARTS~\cite{parts-lilijestrand-sec19} and PTAuth~\cite{ptauth-farkhani-sec21},
relying on the security guarantees provided by the \emph{non-copyability}
property. }
\CH{
%
The security benefits of the non-copyability property are demonstrated by the
PAC reuse attack in the vulnerable code in~\autoref{c:noncopy}. 
PARTS~\cite{parts-lilijestrand-sec19} is vulnerable
to this attack while \sys is not. If two pointers have the same modifier
(type-id in PARTS) and point to the same address, then the processor will
generate the same PAC, and thus they can be used interchangeably at
a different code location. This is possible in PARTS if both pointers have the
same LLVM \emph{ElementType}. This is similar in concept to the COOP attack in
terms of pointer manipulation.
Our incorporation of a pointer location (\cc{\&p}) into the modifier with the
non-copyability property blocks this attack by binding a signed PAC to
a specific pointer location in the code. This binding will not allow a signed
pointer to be used from a different pointer location. 
%
%
\begin{figure}[t!]
\begin{minted}{cpp}
\begin{Verbatim}[commandchars=\\\{\},codes={\catcode`\$=3\catcode`\^=7\catcode`\_=8}]
\PY{n}{T} \PY{n}{foo}\PY{p}{,}\PY{n}{bar}\PY{p}{;} \PY{err}{@}\PY{err}{\PYZbs{}}\PY{n}{label}\PY{p}{\PYZob{}}\PY{n+nl}{c}\PY{p}{:}\PY{n+nl}{noncopy}\PY{p}{:}\PY{n}{l1}\PY{p}{\PYZcb{}}\PY{err}{@}
\PY{n}{foo}\PY{p}{.}\PY{n}{funcptr} \PY{o}{=} \PY{o}{\PYZam{}}\PY{n}{printf}\PY{p}{;} \PY{err}{@}\PY{err}{\PYZbs{}}\PY{n}{label}\PY{p}{\PYZob{}}\PY{n+nl}{c}\PY{p}{:}\PY{n+nl}{noncopy}\PY{p}{:}\PY{n}{l2}\PY{p}{\PYZcb{}}\PY{err}{@}
\PY{n}{bar}\PY{p}{.}\PY{n}{funcptr} \PY{o}{=} \PY{o}{\PYZam{}}\PY{n}{system}\PY{p}{;} \PY{err}{@}\PY{err}{\PYZbs{}}\PY{n}{label}\PY{p}{\PYZob{}}\PY{n+nl}{c}\PY{p}{:}\PY{n+nl}{noncopy}\PY{p}{:}\PY{n}{l3}\PY{p}{\PYZcb{}}\PY{err}{@}
\PY{n}{T} \PY{o}{*}\PY{n}{p} \PY{o}{=} \PY{o}{\PYZam{}}\PY{n}{foo}\PY{p}{;} \PY{c+c1}{// p stores a valid PAC of foo @\PYZbs{}label\PYZob{}c:noncopy:l4\PYZcb{}@}
\PY{n}{T} \PY{o}{*}\PY{n}{q} \PY{o}{=} \PY{o}{\PYZam{}}\PY{n}{bar}\PY{p}{;} \PY{c+c1}{// q stores a valid PAC of bar @\PYZbs{}label\PYZob{}c:noncopy:l5\PYZcb{}@}
\PY{c+c1}{// An attacker performs arbitrary read/write here }
\PY{c+c1}{// (by exploiting a known vulnerability) }
\PY{c+c1}{// to overwrite p as q, i.e., p = q;}
\PY{c+c1}{// now p stores a valid PAC of \PYZam{}bar}
\PY{n}{p}\PY{o}{\PYZhy{}}\PY{o}{\PYZgt{}}\PY{n}{funcptr}\PY{p}{(}\PY{p}{)}\PY{p}{;}  \PY{c+c1}{// Runs system() in PARTS }
               \PY{c+c1}{// because type of p and q are the same @\PYZbs{}label\PYZob{}c:noncopy:l8\PYZcb{}@}
\end{Verbatim}
\end{minted}
	\spacebelowcaption
    \caption{
    Example of vulnerable code that \sys defends against but 
    PARTS~\cite{parts-lilijestrand-sec19} cannot.
    }
	\vspace{-2em}
    \label{c:noncopy}
\end{figure}
}

%

\tsubsection{Performance Evaluation}
\label{s:eval:perf}

\subsubsection{Benchmarks}
\label{s:eval:perf:bench}

\begin{figure*}[t!]
  \centering
  \input{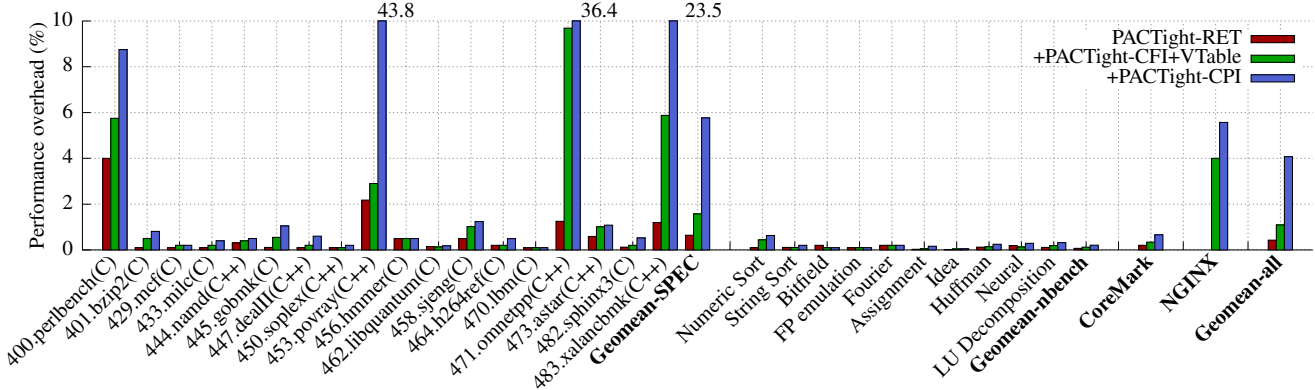}
  \vspace{-1em}	
  \caption{
  The performance overhead of SPEC CPU2006, nbench, and CoreMark relative to an
  unprotected baseline build. Our three \sys protections are: 1) return
  addresses (\sys-RET), 2) CFI, C++ VTable protection and return addresses
  (+\sys-CFI+VTable), and 3) CPI offering full protection for all sensitive
  pointers (+\sys-CPI).
  }
  \vspace{-1.2em}
  \label{f:perf}
\end{figure*}

\begin{figure}[t!]
  \centering
  \input{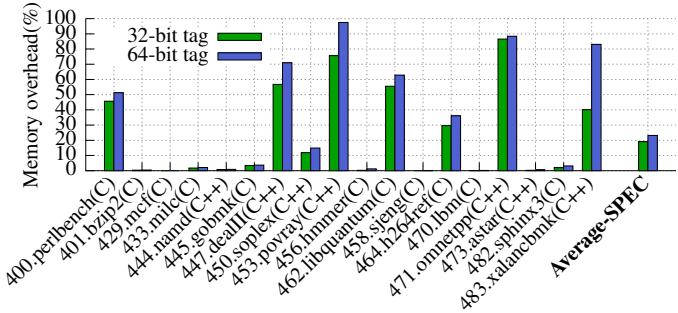}
  \caption{
  The memory overhead of SPEC CPU2006 for \sys-CPI, \sys's highest protection
  mechanism. \sys imposes a low overhead of 23.2\% \CH{and 19.1\%} on
  average \CH{for 64-bit and 32-bit tags, respectively}.
  }
  \vspace{-1.2em}	
  \label{f:mem}
\end{figure}

\begin{figure}[t!]
  \centering
  \input{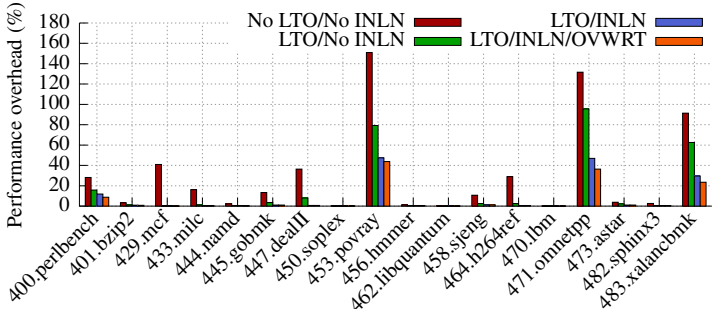}
  \spacebelowcaption
  \caption{
   \textbf{Impact of the performance optimizations.}
   (LTO: Link Time Optimization   
   INLN: inlining APIs;
   OVWRT: overwrite stripped pointer with PACed pointer)}
  \vspace{-1.8em}
  \label{f:opt-impact}
\end{figure}

%

\pp{Benchmark applications.}
For our performance evaluation, we use three benchmarks: SPEC
CPU2006~\cite{spec-cpu2006-web}, nbench~\cite{nbench}, and
CoreMark~\cite{coremark}, 
and one real-world application, NGINX web server~\cite{NGINX-web}.
In order to run the SPEC CPU2006 benchmark suite,
we ported each SPEC benchmark to Apple M1 and
built it from scratch. We were not able to run one benchmark, \cc{403.gcc},
on the Apple M1 even with Apple's vanilla Clang/LLVM compiler.
We suspect a bug in the MacOS/M1 toolchain.
We ran all benchmark applications with real PA instructions except for seven
C++ benchmarks in the SPEC benchmark. For the C++ benchmarks, we replaced a PA
instruction with seven \cc{eor} instructions to emulate the overhead of the PA
instructions as discussed in~\autoref{s:eval:method}.

\pp{Performance overhead.}
\autoref{f:perf} shows the performance of the \sys defenses on the individual SPEC
benchmarks, nbench, and CoreMark. The SPEC benchmarks have a geometric mean of
0.64\%, 1.57\%, and \CH{5.77\%} for \sys-RET, \sys-CFI+VTable+RET, and \sys-CPI,
respectively. The geometric means of all benchmark applications are 0.43\%,
1.09\%, and \CH{4.07\%} for \sys-RET, \sys-CFI+VTable+RET, and \sys-CPI,
respectively. As can be seen, \sys has very low overhead on almost all
benchmarks and across all the protection mechanisms.
The exceptions here are \cc{453.povray}, \cc{471.omnetpp}, and
\cc{483.xalancbmk} for \sys-CPI. \CH{We discuss these further 
in~\autoref{s:appendix:anal}.}


We evaluated NGINX on the Apple M1 using its 4 big cores to
stress the machine. We used the same configuration 
used to bench NGINX TLS transactions per second~\cite{nginx:ssltest}. We used
the HTTP benchmarking tool \cc{wrk}~\cite{github:wrk} to generate concurrent
HTTP requests. 
We ran \cc{wrk} on another machine under the same network. Each \cc{wrk}
spawns three threads where each thread handles 50 connections. We
observed a small performance overhead: 4\% for \sys-CFI and 5.57\% for \sys-CPI.


\pp{Memory overhead.}
In order to see how much additional memory is used by \sys's metadata store, we
measured the maximum resident set size (RSS) during the execution of the SPEC
CPU2006 benchmarks. We ran the SPEC benchmarks with the \sys-CPI protection
because it is the highest level of protection in \sys, thus it requires the
largest number of entries in the metadata store. \CH{We used both 64-bit and
32-bit tag sizes to measure the gain if we used a smaller tag. The size of the
metadata is 16 bytes in the case of a 64-bit tag, and 12 bytes in the case of
a 32-bit tag.} \autoref{f:mem} shows the results of our measurements. In spite
of measuring the highest security mechanism with the most instrumentations,
\sys imposes an overhead of 23\% on average \CH{for 64-bit tags and 19\% on
average for 32-bit tags. The memory overhead is proportional to $O(n)$ where
$n$ is the number of sensitive pointers, with the metadata size being either
2\x the size of the pointer (64-bit tag) or 1.5\x the size of the pointer
(32-bit tag).}

\subsubsection{\CH{Impact of Optimizations}}
\label{s:eval:perf:opt}

\CH{ Here we showcase the impact of the optimizations discussed in
\autoref{s:defense:opt} and \autoref{s:impl}. We added three optimizations to
\sys to improve performance: Link Time Optimization (LTO), inlining of \sys
runtime library functions (INLN), and overwriting the stripped pointer with
a PACed pointer (OVWRT). \autoref{f:opt-impact} shows the performance overhead
of \sys-CPI with and without the optimizations in various configurations. As
can be seen, the optimizations were critical to greatly improving \sys's performance.
}

\section{Discussion and Limitations}
\label{s:discussion}


\pp{Information leakage attack on the metadata store.}
In our threat model, 
an attacker is able to access the \sys metadata store while it is
probabilistically hidden using address space layout randomization (ASLR).
%
Even if the \sys metadata is leaked, an attacker is not able to
exploit the leaked information. In order for an attacker to take advantage of
the leakage, she has to launch an attack from a different location and this is
already protected by the non-copyability property. \CH{The only part of the 
modifier that gets leaked is the random tag, but the location (\cc{\&p}) in the modifier
still enforces the non-copyability property.}
\CH{In regards to legal and illegal pointers, \sys always authenticates the right hand 
side of a PACTight-signed pointer assignment. Thus, if a pointer in the right hand 
side is illegal, its authentication will fail. In this way, \sys prevents the 
propagation of illegal pointers.}
Another hypothetical case is that an attacker reuses a random tag to
bypass the non-dangling property.
While such an attack is possible in theory, the bar is very high in practice. An
attacker cannot reuse dangling pointers at an arbitrary location due to the
non-copyability property, and this significantly limits the attack.
%
%
%
Moreover, we argue this is not a fundamental flaw in \sys's design. The random
tag can be enforced using ARM's new Memory Tagging Extension (MTE) 
feature~\cite{arm:mte}.
The presence of MTE will mitigate the random tag reuse attacks since the tags
are protected in physical memory that can never be accessed by an attacker.
\sys can easily be extended to utilize MTE as a tag store.

\section{Related Work}
\label{s:relwk}

\CH{In this section, we only discuss related studies that
have not been discussed previously .}

\pp{Cryptographic pointer defenses.}
CCFI~\cite{mashtizadeh:ccfi} uses MACs to protect return addresses, function
pointers, and VTable pointers. 
Conceptually, the use of MACs is similar to PA. But, since CCFI does not benefit 
from the hardware-accelerated PA instructions, it has an average of 52\% overhead 
across SPEC CPU2006 benchmarks.

\pp{Integrity policies.}
Control-Flow Integrity (CFI)~\cite{cfi} restricts the valid target sites for
indirect control-flow transfers. Static CFI schemes are vulnerable to
control-flow bending~\cite{carlini:cfb}. 
Since \sys-CFI-VTable seals a pointer
with its location and a random tag, this limits the feasibility of a reuse
attack. Other 
dynamic approaches require additional threads to analyze
data from Intel Processor Trace~\cite{liu:flowguard, gu:pt-cfi, ge:griffin,
hu:ucfi} limiting scalability.

Code Pointer Integrity (CPI)~\cite{kuznetsov:cpi} protects sensitive pointers
(code pointers and pointers that refer to code pointers) by storing the
sensitive pointers in a seperate hidden memory region. Return addresses are
stored on a safe stack. \sys-CPI provides temporal safety to sensitive
pointers, which CPI does not, and protects virtual function pointers in addition
to sensitive pointers, all while having a lower overhead across all defenses.

\CH{CFIXX~\cite{burow:cfixx} protects VTable pointers by enforcing Object Type
Integrity (OTI). CFIXX stores metadata on construction and
checks the metadata at the virtual function call site. CFIXX incurs an 
overhead of 4.98\%. \sys-CFI+VTable incurs lower overhead (1.98\%) whilst
providing stronger guarantees by enforcing CFI.}

%
%
%
%

\pp{Temporal memory safety.}
Explicit pointer invalidation is a common strategy to enforce temporal memory
safety. DangNull~\cite{lee:dangnull}, DangSan~\cite{dangsan-kouwe-eurosys17},
FreeSentry~\cite{freesentry-younan-ndss15}, pSweeper~\cite{psweeper-liu-ccs18},
and BOGO~\cite{bogo-zhang-asplos19} invalidate all pointers to an object when
the object is freed. These schemes typically incur high costs.
CRCount~\cite{crcount-shin-ndss19} implicitly invalidates pointers by using
reference counting. This approach comes at memory costs since some objects may
never be freed. CETS~\cite{nagarakatte:cets} uses disjoint metadata to check if
an object still exists upon pointer dereferences.
\CH{MarkUs~\cite{markus-ainsworth-sp20} is a memory allocator that protects
from use-after-free attacks. It quarantines freed data and prevents
reallocation until there are no dangling pointers. In contrast, \sys
offers broader protection and protects sensitive pointers from memory
corruption attacks.}

\tsection{Conclusion}
\label{s:conclusion}

We presented \sys, an efficient and robust mechanism to guarantee pointer
integrity using ARM's Pointer Authentication mechanism. We identified three
security properties \sys enforces to ensure pointer integrity: (1)
Unforgeability: a pointer cannot be forged to point to an unintended memory
object. (2) Non-copyability: a pointer cannot be copied and re-used
maliciously. (3) Non-dangling: a pointer cannot refer to an unintended memory
object if the object has been freed. We implememented \sys with four defense
mechanisms, protecting forward edge, backward edge, virtual function pointers, and 
sensitive pointers. We demonstrated the security of \sys against real
and synthesized attacks and showcased its low performance and memory overhead,
\CH{4.07\%} and 23.2\%, on average respectively,
using real PAC instructions.

\section*{Acknowledgment}
\label{s:ack}

This work is supported in part by the U.S. Office of Naval Research under grants
N00014-18-1-2022, the U.S. Nuclear Regulatory Commission under grants
31310021M0005, and the Federal Aviation Administration under grants
A51-A11L.UAS.92.

\onecolumn
\begin{multicols}{2}
\bibliographystyle{plainurl}
\footnotesize
\bibliography{p,cfi,sslab,pactight,conf}
\end{multicols}


\clearpage
\setcounter{section}{0}
\setcounter{page}{1}
\renewcommand{\thepage}{A-\arabic{page}}
\renewcommand{\thesection}{A.\arabic{section}}
\newpage
\twocolumn
\section*{\CH{Appendix}}

\CH{We describe the \sys runtime library functions in detail
(\autoref{s:appendix:lib}), its collision likelihood
(\autoref{s:appendix:collision}), showcase \sys instrumentation statistics
(\autoref{s:appendix:stat}), and finally analyze the high overhead benchmarks
(\autoref{s:appendix:anal}).}


\section{\CH{\sys Runtime Library Functions}}
\label{s:appendix:lib}

\CH{\autoref{c:libpac} presents the (simplified) \sys library function code.
The two omitted functions, \cc{meta\_hashtable\_get} and
\cc{meta\_hashtable\_set}, retrieve and set the metadata, respectively.}

\begin{figure}[h!]
\begin{minted}{cpp}
\begin{Verbatim}[commandchars=\\\{\},codes={\catcode`\$=3\catcode`\^=7\catcode`\_=8}]
\PY{c+cm}{/** == Pseudocode for the library}
\PY{c+cm}{ * functions in PACTight== **/}

\PY{n}{pct\PYZus{}add\PYZus{}tag}\PY{p}{(}\PY{k+kt}{void}\PY{o}{*} \PY{n}{p}\PY{p}{,} \PY{k+kt}{int} \PY{n}{tsz}\PY{p}{,} \PY{k+kt}{int} \PY{n}{asz}\PY{p}{)}\PY{p}{\PYZob{}}
   \PY{n}{tag} \PY{o}{=} \PY{n}{CSPRNG}\PY{p}{(}\PY{p}{)}\PY{p}{;} \PY{c+c1}{// Generate random tag}
   \PY{k}{for}\PY{p}{(}\PY{k+kt}{int} \PY{n}{i} \PY{o}{=} \PY{l+m+mi}{0}\PY{p}{;} \PY{n}{i} \PY{o}{\PYZlt{}} \PY{n}{asz}\PY{p}{;} \PY{n}{i}\PY{o}{+}\PY{o}{+}\PY{p}{)}\PY{p}{\PYZob{}}
      \PY{n}{meta\PYZus{}hashtable\PYZus{}set}\PY{p}{(}\PY{n}{p}\PY{p}{,} \PY{n}{tag}\PY{p}{,} \PY{n}{tsz}\PY{p}{,} \PY{n}{asz} \PY{o}{\PYZhy{}} \PY{n}{i}\PY{p}{)}\PY{p}{;}
      \PY{n}{p} \PY{o}{=} \PY{n}{p} \PY{o}{+} \PY{n}{tsz}\PY{p}{;}
   \PY{p}{\PYZcb{}}
\PY{p}{\PYZcb{}}

\PY{n}{pct\PYZus{}sign}\PY{p}{(}\PY{k+kt}{void}\PY{o}{*}\PY{o}{*} \PY{n}{p}\PY{p}{)}\PY{p}{\PYZob{}}
   \PY{n}{metadata}\PY{o}{*} \PY{n}{meta} \PY{o}{=} \PY{n}{meta\PYZus{}hashtable\PYZus{}get}\PY{p}{(}\PY{o}{*}\PY{n}{p}\PY{p}{)}\PY{p}{;}
   \PY{n}{pac\PYZus{}modifier} \PY{o}{=} \PY{n}{p} \PY{o}{\PYZca{}} \PY{n}{meta}\PY{o}{\PYZhy{}}\PY{o}{\PYZgt{}}\PY{n}{tag}\PY{p}{;}
   \PY{k+kr}{\PYZus{}\PYZus{}asm} \PY{n+nf}{volatile} \PY{p}{(}\PY{l+s}{\PYZdq{}}\PY{l+s}{pacia \PYZpc{}x[pointer], \PYZpc{}x[modifier]}\PY{l+s+se}{\PYZbs{}n}\PY{l+s}{\PYZdq{}}
           \PY{o}{:} \PY{p}{[}\PY{n}{pointer}\PY{p}{]} \PY{l+s}{\PYZdq{}}\PY{l+s}{+r}\PY{l+s}{\PYZdq{}} \PY{p}{(}\PY{o}{*}\PY{n}{p}\PY{p}{)}
           \PY{o}{:} \PY{p}{[}\PY{n}{modifier}\PY{p}{]} \PY{l+s}{\PYZdq{}}\PY{l+s}{r}\PY{l+s}{\PYZdq{}}\PY{p}{(}\PY{n}{pac\PYZus{}modifier}\PY{p}{)}
              \PY{p}{)}\PY{p}{;}
\PY{p}{\PYZcb{}}

\PY{n}{pct\PYZus{}auth}\PY{p}{(}\PY{k+kt}{void}\PY{o}{*}\PY{o}{*} \PY{n}{p}\PY{p}{,} \PY{k+kt}{void}\PY{o}{*} \PY{n}{p\PYZus{}index}\PY{p}{)}\PY{p}{\PYZob{}}
   \PY{n}{metadata}\PY{o}{*} \PY{n}{meta} \PY{o}{=} \PY{n}{meta\PYZus{}hashtable\PYZus{}get}\PY{p}{(}\PY{n}{p\PYZus{}index}\PY{p}{)}\PY{p}{;}
   \PY{n}{pac\PYZus{}modifier} \PY{o}{=} \PY{n}{p} \PY{o}{\PYZca{}} \PY{n}{meta}\PY{o}{\PYZhy{}}\PY{o}{\PYZgt{}}\PY{n}{tag}\PY{p}{;}
   \PY{k+kr}{\PYZus{}\PYZus{}asm} \PY{n+nf}{volatile} \PY{p}{(}\PY{l+s}{\PYZdq{}}\PY{l+s}{autia \PYZpc{}x[pointer], \PYZpc{}x[modifier]}\PY{l+s+se}{\PYZbs{}n}\PY{l+s}{\PYZdq{}}
           \PY{o}{:} \PY{p}{[}\PY{n}{pointer}\PY{p}{]} \PY{l+s}{\PYZdq{}}\PY{l+s}{+r}\PY{l+s}{\PYZdq{}} \PY{p}{(}\PY{o}{*}\PY{n}{p}\PY{p}{)}
           \PY{o}{:} \PY{p}{[}\PY{n}{modifier}\PY{p}{]} \PY{l+s}{\PYZdq{}}\PY{l+s}{r}\PY{l+s}{\PYZdq{}}\PY{p}{(}\PY{n}{pac\PYZus{}modifier}\PY{p}{)}
           \PY{p}{)}\PY{p}{;}
\PY{p}{\PYZcb{}}

\PY{n}{pct\PYZus{}rm\PYZus{}tag}\PY{p}{(}\PY{k+kt}{void}\PY{o}{*} \PY{n}{p}\PY{p}{)}\PY{p}{\PYZob{}}
   \PY{n}{metadata}\PY{o}{*} \PY{n}{meta} \PY{o}{=} \PY{n}{meta\PYZus{}hashtable\PYZus{}get}\PY{p}{(}\PY{n}{p}\PY{p}{)}\PY{p}{;}
   \PY{k+kt}{int} \PY{n}{asz} \PY{o}{=} \PY{n}{meta}\PY{o}{\PYZhy{}}\PY{o}{\PYZgt{}}\PY{n}{asz}\PY{p}{;}
   \PY{k+kt}{int} \PY{n}{tsz} \PY{o}{=} \PY{n}{meta}\PY{o}{\PYZhy{}}\PY{o}{\PYZgt{}}\PY{n}{tsz}\PY{p}{;}
   \PY{k}{for}\PY{p}{(}\PY{k+kt}{int} \PY{n}{i} \PY{o}{=} \PY{l+m+mi}{0}\PY{p}{;} \PY{n}{i} \PY{o}{\PYZlt{}} \PY{n}{asz}\PY{p}{;} \PY{n}{i}\PY{o}{+}\PY{o}{+}\PY{p}{)}\PY{p}{\PYZob{}}
      \PY{n}{meta\PYZus{}hashtable\PYZus{}remove}\PY{p}{(}\PY{n}{p}\PY{p}{)}\PY{p}{;}
      \PY{n}{p} \PY{o}{=} \PY{n}{p} \PY{o}{+} \PY{n}{tsz}\PY{p}{;}
   \PY{p}{\PYZcb{}}
\PY{p}{\PYZcb{}}
\end{Verbatim}
\end{minted}
	\vspace{-10px}
	\caption{
		\CH{\sys runtime library functions.}
	}
	\label{c:libpac}
\end{figure}




\newcommand\tab[1][1cm]{\hspace*{#1}}
\vspace{3mm}


\begingroup
\addtolength{\tabcolsep}{-3pt}
\renewcommand{\arraystretch}{1} 
\begin{table*}[t]
        \centering
        \footnotesize
	\resizebox{\textwidth}{!}{\CH{
\begin{tabular}{|l|r|r|r|r|r|r|r|r|r|r|r|r|r|r|r|r|}
\hline
\multicolumn{1}{|c|}{\textbf{\begin{tabular}[c]{@{}c@{}}Benchmark \\ Name\end{tabular}}} &
  \multicolumn{3}{l|}{\textbf{Compilation time}} &
  \multicolumn{3}{l|}{\textbf{Binary size}} &
  \multicolumn{1}{l|}{\textbf{\begin{tabular}[c]{@{}l@{}}Number \\ of stores\end{tabular}}} &
  \multicolumn{1}{l|}{\textbf{\begin{tabular}[c]{@{}l@{}}Number of \\ protected \\ stores\end{tabular}}} &
  \multicolumn{1}{l|}{\textbf{\begin{tabular}[c]{@{}l@{}}Percentage \\ of protected \\ stores\end{tabular}}} &
  \multicolumn{1}{l|}{\textbf{\begin{tabular}[c]{@{}l@{}}Number \\ of loads\end{tabular}}} &
  \multicolumn{1}{l|}{\textbf{\begin{tabular}[c]{@{}l@{}}Number \\ of protected\\ loads\end{tabular}}} &
  \multicolumn{1}{l|}{\textbf{\begin{tabular}[c]{@{}l@{}}Percentage \\ of protected \\ loads\end{tabular}}} &
  \multicolumn{1}{l|}{\textbf{\begin{tabular}[c]{@{}l@{}}Number of \\ pct\_add\_tag\end{tabular}}} &
  \multicolumn{1}{l|}{\textbf{\begin{tabular}[c]{@{}l@{}}Number of \\ pct\_sign\end{tabular}}} &
  \multicolumn{1}{l|}{\textbf{\begin{tabular}[c]{@{}l@{}}Number of \\ pct\_auth\end{tabular}}} &
  \multicolumn{1}{l|}{\textbf{\begin{tabular}[c]{@{}l@{}}Number of \\ pct\_rm\_tag\end{tabular}}} \\ \hline
 &
  \multicolumn{1}{l|}{\textbf{Vanilla}} &
  \multicolumn{1}{l|}{\textbf{PACTight}} &
  \multicolumn{1}{l|}{\textbf{Overhead}} &
  \multicolumn{1}{l|}{\textbf{Vanilla}} &
  \multicolumn{1}{l|}{\textbf{PACTight}} &
  \multicolumn{1}{l|}{\textbf{Overhead}} &
  \multicolumn{1}{l|}{} &
  \multicolumn{1}{l|}{} &
  \multicolumn{1}{l|}{} &
  \multicolumn{1}{l|}{} &
  \multicolumn{1}{l|}{} &
  \multicolumn{1}{l|}{} &
  \multicolumn{1}{l|}{} &
  \multicolumn{1}{l|}{} &
  \multicolumn{1}{l|}{} &
  \multicolumn{1}{l|}{} \\ \hline
\textbf{400.perlbench (c)}   & 318  & 366  & 15.09\% & 2616  & 2784  & 6.42\%   & 15826 & 1083  & 6.84\%  & 42753  & 9068  & 21.21\% & 310   & 1079  & 9064  & 46   \\ \hline
\textbf{401.bzip2 (c)}       & 47   & 47   & 0.00\%  & 208   & 240   & 15.38\%  & 1754  & 11    & 0.63\%  & 2602   & 150   & 5.76\%  & 4     & 45    & 150   & 2    \\ \hline
\textbf{429.mcf (c)}         & 25   & 25   & 0.00\%  & 104   & 104   & 0.00\%   & 252   & 0     & 0.00\%  & 399    & 0     & 0.00\%  & 0     & 0     & 0     & 0    \\ \hline
\textbf{433.milc (c)}        & 120  & 122  & 1.67\%  & 288   & 288   & 0.00\%   & 889   & 16    & 1.80\%  & 3201   & 35    & 1.09\%  & 2     & 7     & 35    & 2    \\ \hline
\textbf{444.namd (c++)}      & 114  & 115  & 0.88\%  & 424   & 504   & 18.87\%  & 2333  & 53    & 2.27\%  & 6170   & 21    & 0.34\%  & 41    & 64    & 21    & 35   \\ \hline
\textbf{445.gobmk (c)}       & 286  & 297  & 3.70\% & 7696  & 8600  & 10.51\% & 4584  & 6     & 0.13\%  & 17370  & 74    & 0.43\%  & 8     & 10    & 74    & 6    \\ \hline
\textbf{447.dealII (c++)}    & 1508 & 1516 & 0.53\%  & 1488  & 1768  & 18.82\%  & 41257 & 6204  & 15.04\% & 94791  & 8543  & 9.01\%  & 2892  & 6745  & 8036  & 2867 \\ \hline
\textbf{450.soplex (c++)}    & 408  & 434  & 6.37\%  & 840   & 1072  & 27.62\%  & 5409  & 254   & 4.70\%  & 16665  & 724   & 4.34\%  & 242   & 632   & 441   & 52   \\ \hline
\textbf{453.povray (c++)}    & 625  & 653  & 4.48\%  & 2496  & 3120  & 25.00\%  & 15128 & 474   & 3.13\%  & 25766  & 2247  & 8.72\%  & 117   & 525   & 2029  & 36   \\ \hline
\textbf{456.hmmer (c)}       & 141  & 148  & 4.96\%  & 384   & 456   & 18.75\%  & 3618  & 33    & 0.91\%  & 8557   & 264   & 3.09\%  & 16    & 28    & 264   & 16   \\ \hline
\textbf{462.libquantum (c)}  & 32   & 33   & 3.13\%  & 144   & 144   & 0.00\%   & 270   & 0     & 0.00\%  & 585    & 0     & 0.00\%  & 0     & 0     & 0     & 0    \\ \hline
\textbf{458.sjeng (c)}       & 61   & 62   & 1.61\% & 368   & 368   & 0.00\%   & 1899  & 0     & 0.00\%  & 3570   & 1     & 0.03\%  & 1     & 1     & 1     & 1    \\ \hline
\textbf{464.h264ref (c)}     & 296  & 304  & 2.70\%  & 1248  & 1568  & 25.64\%  & 11309 & 88    & 0.78\%  & 27103  & 1659  & 6.12\%  & 48    & 139   & 1663  & 11   \\ \hline
\textbf{470.lbm (c)}         & 12   & 13   & 8.33\%  & 104   & 104   & 0.00\%   & 99    & 0     & 0.00\%  & 269    & 0     & 0.00\%  & 0     & 0     & 0     & 0    \\ \hline
\textbf{471.omnetpp (c++)}   & 567  & 570  & 0.53\%  & 2136  & 2528  & 18.35\%  & 6007  & 1158  & 19.28\% & 8697   & 2890  & 33.23\% & 264   & 1206  & 2025  & 66   \\ \hline
\textbf{473.astar (c++)}     & 34   & 35   & 2.86\% & 144   & 144   & 0.00\%   & 708   & 0     & 0.00\%  & 1191   & 2     & 0.17\%  & 0     & 0     & 1     & 0    \\ \hline
\textbf{482.sphinx3 (c)}     & 109  & 110  & 0.92\%  & 384   & 416   & 8.33\%   & 1421  & 20    & 1.41\%  & 4716   & 152   & 3.22\%  & 2     & 18    & 152   & 2    \\ \hline
\textbf{483.xalancbmk (c++)} & 3149 & 3624 & 15.08\% & 11184 & 19800 & 77.04\%  & 39741 & 10834 & 27.26\% & 110595 & 35025 & 31.67\% & 3820  & 13207 & 33046 & 1065 \\ \hline
\textbf{Average/Total} &
  \multicolumn{1}{l|}{} &
  \multicolumn{1}{l|}{} &
  \textbf{3.14\%} &
  \multicolumn{1}{l|}{} &
  \multicolumn{1}{l|}{} &
  \textbf{13.87\%} &
  \textbf{152504} &
  \textbf{20234} &
  \textbf{13.27\%} &
  \textbf{375000} &
  \textbf{60855} &
  \textbf{16.23\%} &
  \textbf{7767} &
  \textbf{23706} &
  \textbf{57002} &
  \textbf{4207} \\ \hline
\end{tabular}
}
}
	\caption{\CH{Instrumentation statistics for \sys-CPI in SPEC CPU2006.}
        %
        %
    }
        \label{t:stat}
\end{table*}

%

%

\section{\CH{Collision Likelihood of \sys}}
\label{s:appendix:collision}


\CH{ A determined attacker can attempt to break our \sys scheme using modifier
collisions. For example, if an attacker allocates a PAC'd pointer \cc{p} with
random tag \cc{A} at location \cc{B}, then deallocates and reallocates \cc{p}
with a different random tag \cc{C}. Then, the attacker can reuse \cc{p} at
a different location \cc{D} if (\cc{A} XOR \cc{B}) collides with (\cc{C} XOR
\cc{D}). The probability that this collision occurs is extremely low. Because
we XOR the modifier with a 64-bit random tag, the distribution of PAC modifiers
is uniformly random with 64 bits of entropy (\ie, $2^{64}$); therefore, an
attacker cannot practically break the non-copyability property via modifier
collisions.}

\CH{ Alternatively, an attacker can attempt to break the scheme using PAC
collisions. By reusing a sensitive pointer at many different locations, there
is a chance (albeit a very low probability) that the same PAC could be
generated even though the modifiers are different. This would require an
expected $2^{b}$ guesses, where $b$ is the number of bits available for PAC
($b=16$ in ARMv8.3-A). The birthday problem~\cite{crypto-simple} does not apply
in this case since an attacker has no way to efficiently bruteforce many
pointers at the same time. This attack is equivalent to attempting to forge an
authenticated pointer. Consequently, these collision attacks are not feasible
against the metadata scheme that \sys proposes.}

\section{\CH{Instrumentation Statistics}}
\label{s:appendix:stat}

\CH{\autoref{t:stat} shows various instrumentation statistics for \sys-CPI in SPEC
CPU2006. These include compilation time, binary size, the total and protected
number of loads and stores, and the number of instrumentations of \pctaddtag,
\pctsign, \pctauth and \pctrmtag. As shown, \sys-CPI imposes a marginal
overhead in compilation time increase and binary size increase, 3.14\% and
13.87\%, respectively. For some benchmarks, the overhead is not directly
proportional to the number of instrumentations. This is because the
instrumentations may be called several times in a loop for example.} 

\CH{There are a few of these benchmarks that show zero instrumentation. We
investigated this and found that the compiler optimizes out the sensitive
\cc{load} and \cc{store} instructions. Running \sys without compiler
optimizations produces the instrumentations accordingly. Thus, there are no
false negatives with these benchmarks.}
%

\section{\CH{Analysis on High Overhead Benchmarks}}
\label{s:appendix:anal}

\CH{Three benchmarks in SPEC CPU2006, namely \cc{453.povray}, \cc{471.omnetpp} and
\cc{483.xalancbmk}, have high performance overhead with \sys-CPI than the rest
of the benchmarks. In this section, we analyze why these benchmarks have higher
overhead and suggest possible improvements.}



\pp{\cc{\CH{453.povray}}.}
\CH{The overhead is mainly due to the loops using sensitive data pointers inside.
Specifically, the \cc{struct Method\_struct} is one of the struct types in
\cc{453.povray} that is considered to be sensitive. This struct has a series of
function pointers and is used like a virtual function table. Pointers of type
\cc{struct Method\_struct} and its members are used in loop conditions and
inside loops. Since these pointers are sensitive, \sys-CPI enforces protection
on them. This means calling \pctauth when they are dereferenced and re-adding
the PAC with \pctsign. This happens multiple times in one loop iteration,
causing extra overhead shown in~\autoref{f:perf}.}
%

%

\pp{\CH{\cc{471.omnetpp} and \cc{483.xalancbmk}}.}
\CH{The main source of the overhead is frequent virtual function call, much more
than other C++ benchmarks. For every virtual function, \sys-CPI calls \pctauth
to authenticate the virtual function pointer and \pctsign to re-add the PAC.}

\pp{\CH{Comparison to prior work.}}
\CH{\sys-CPI's overhead in these three benchmarks is similar to the original
CPI~\cite{kuznetsov:cpi}. Note that \sys-CPI has double the instrumentation,
since it needs to authenticate and then re-sign, whilst CPI would only compare
with its metadata with a single instrumentation.
PARTS~\cite{parts-lilijestrand-sec19} does not evaluate SPEC CPU2006. Even
though PTAuth~\cite{ptauth-farkhani-sec21} does evaluate with a subset of the
SPEC CPU2006 benchmarks, they do not mention the performance numbers for
cc{453.povray}, \cc{471.omnetpp} and \cc{483.xalancbmk}. We expect that these
numbers would be quite high. PACStack~\cite{pacstack-lilijestrand-sec21}
evaluates with SPEC CPU2017. Thus, we could only compare definitively with the
original CPI.}

\pp{\CH{Possible improvement}}
\CH{One of the main reasons overhead is very high is because of instrumentation inside loops.
One could say that a possible solution would be hoisting the
authenticating and signing to be outside loops, \ie, authenticate before
entering a loop and sign after the loop is done. In that way, authenticating and
signing would only be done once.}

\end{document}